\documentclass[a4paper,11pt]{article}

\usepackage[utf8]{inputenc}
\usepackage[T1]{fontenc}

\usepackage{jheppub}

\usepackage{doi}%<----------
\usepackage{hyperref}
\hypersetup{
  colorlinks=true,        % false: boxed links; true: colored links
  linkcolor=magenta,      % color of internal links
  citecolor=red,         % color of links to bibliography
}

\usepackage{graphicx}% Include figure files
\usepackage{dcolumn}% Align table columns on decimal point
\usepackage{bm}% bold math
\usepackage{color}
\usepackage{enumitem}
\usepackage{amsmath}
\usepackage{amssymb}

\newcommand{\beq}{\begin{equation}}
\newcommand{\eeq}{\end{equation}}
\newcommand{\bea}{\begin{eqnarray}}
\newcommand{\eea}{\end{eqnarray}}

\usepackage{bbm}
\usepackage{amsfonts}
\usepackage{mathrsfs}
\usepackage{latexsym}
\usepackage{epsfig}
\usepackage{epstopdf}
\usepackage{epstopdf}
\usepackage{graphicx}
\usepackage{amssymb}
\usepackage{amsmath}
\usepackage{dcolumn}
\usepackage{bm}
\usepackage{color}
\usepackage{comment}
\usepackage{xcolor}

\title{\bf Accretion flow around Kerr metric in the infra-red limit of asymptotically safe gravity}

\author[a]{Orhan~Donmez}
\emailAdd{orhan.donmez@aum.edu.kw}
\affiliation[a]{College of Engineering and Technology, American University of the Middle East, Egaila 54200, Kuwait}

\author[b,c]{Sushant G. Ghosh }
\emailAdd{sghosh2@jmi.ac.in}
\affiliation[b]{Centre for Theoretical Physics, Jamia Millia Islamia, New Delhi 110025, India}
\affiliation[c]{Astrophysics and Cosmology Research Unit, School of Mathematics, Statistics and Computer Science, University of KwaZulu-Natal, Private Bag 54001, Durban 4000, South Africa}

\author[d]{M. Yousaf}
\emailAdd{myousaf.math@gmail.com}
\affiliation[d]{Department of Mathematics, Virtual University of Pakistan, 54-Lawrence Road, Lahore 54000, Pakistan}

\author[e,f]{G. Mustafa}
\emailAdd{gmustafa3828@gmail.com}
\affiliation[e]{Department of Physics, Zhejiang Normal University, Jinhua 321004, China}
\affiliation[f]{Research Center of Astrophysics and Cosmology, Khazar University, 41 Mehseti Street, AZ1096, Baku, Azerbaijan}

\author[g,h,i]{Farruh~Atamurotov}
\emailAdd{atamurotov@yahoo.com}
\affiliation[g]{University of Tashkent for Applied Sciences, Str. Gavhar 1, Tashkent 100149, Uzbekistan}
\affiliation[h]{Tashkent State Technical University, Tashkent 100095, Uzbekistan}
\affiliation[i]{Urgench State University, Kh. Alimdjan str. 14, Urgench 220100, Uzbekistan}

\abstract{
We investigate accretion disk dynamics and the formation of quasi-periodic oscillations (QPOs) in the infrared limit around Kerr-like black holes in asymptotically safe gravity. Relativistic hydrodynamic solutions of Bondi-Hoyle-Lyttleton (BHL) accretion reveal that quantum corrections significantly modify the structure of the shock cone formed around the black hole. The black hole spin controls the azimuthal asymmetry of the shock cone through frame-dragging effects, whereas the quantum correction parameter effectively reduces the strength of gravitational focusing by modifying the metric coefficients in the strong-field region, resulting in a wider shock opening angle, weaker post-shock compression, and reduced density concentration within the cone. Time-dependent mass accretion rates reveal oscillation modes trapped within the shock cone. The power spectral density (PSD) investigations suggest that these modes naturally generate low-frequency QPOs, whose amplitudes, coherence, and harmonic structure depend on both the spin and the quantum correction parameter. The PSD analyses performed at different radial locations reveal that identical QPO frequencies are obtained in all cases. The numerically detected frequencies result from the excitation of global oscillation modes trapped within the post-shock region. The resulting global modes are found to consist of fundamental frequencies, their associated harmonic overtones, and near-commensurate frequency ratios such as 2:1 and 3:2. Coherent oscillations are enhanced and near-commensurate frequency ratios are produced when moderate rotation and moderate quantum corrections are coupled. Large quantum correction parameters, on the other hand, wash out unique spectral peaks and suppress oscillation amplitudes.
}


\begin{document} 
\maketitle
\flushbottom

\section{Introduction}

Einstein’s general theory of relativity (GTR) achieved remarkable success in explaining a broad spectrum of classical and astrophysical phenomena. One of the most profound open questions in theoretical physics is the search for a viable quantum description of gravitation. Whereas GTR exhibits intrinsic limitations when applied to extreme physical regimes, such as gravitational interactions probing distances comparable to the Planck length, spacetime singularities, and ultra-strong gravitational fields, classical predictions are expected to break down, necessitating a deeper microscopic theory \cite{penrose1965gravitational,hawking1970proceedings}. In parallel, a wealth of observational evidence established that the Universe is presently undergoing an accelerated expansion \cite{riess1998observational,perlmutter1999measurements,riess2007new,nojiri2007introduction,Bamba2012}. These observations imply that nearly $95\%$ of the total cosmic energy budget is composed of nonluminous constituents, namely dark matter (DM) and dark energy (DE) \cite{peebles2003cosmological,khlopov2014dark,narzilloev2021dynamics}, whereas DE, characterized by an effective negative pressure, is widely regarded as the driving mechanism behind the observed cosmic acceleration \cite{nojiri2007introduction,bamba2014cosmology,olmo2019stellar}. Moreover, the singularity theorems emphasised fundamental constraints on the predictivity of GTR, motivating the development of alternative gravitational frameworks capable of addressing these pathologies \cite{rovelli2004quantum,brunnemann2006cosmological,bodendorfer2013new,long2019coherent,bojowald2001absence}. 

Despite decades of intensive research, a complete and universally accepted framework has yet to emerge. Several fundamentally different strategies were proposed, most notably loop quantum gravity \cite{rovelli2008loop,chiou2015loop}. Among these extensions and modifications of GTR proposed, Einstein-Cartan theory incorporates spacetime torsion as a natural geometric extension \cite{hehl1976general}, while string theory and loop quantum gravity introduce fundamentally new approaches to spacetime quantization, offering potential resolutions of singular behavior \cite{green1983superstring,ashtekar2011loop,gambini2008black,ALBADAWI2025102206}. Additionally, some developments in super Yang-Mills theories further extend the applicability of the Einstein framework to study asymptotic symmetries and black hole (BH) physics \cite{koutrolikos2024just,barnich2013einstein}, while some other phenomenological extensions of GTR are in the form of modified gravity models \cite{jackiw1976vacuum,brink1977supersymmetric,harko2010f,harko2011f,nojiri2017modified,dorigoni2024electromagnetic}. 

AA compelling alternative to perturbative renormalisation proposed by Weinberg in the form of the asymptotic safety scenario \cite{Weinberg1979general}, which is based on the possibility that gravity admits a nontrivial fixed point under the renormalisation group flow, ensuring that all essential couplings remain finite at arbitrarily high energies. Gravity can be consistently described as a quantum field theory at all scales without the addition of new degrees of freedom if such a fixed point exists. Extensive analytical and numerical studies, detailed in \cite{niedermaier2006asymptotic}, were inspired by the premise that gravitational interactions meet the requirements for asymptotic safe gravity.

The possible phenomenological implications of asymptotically safe gravity are an essential feature. Compact astrophysical objects, in particular, offer a perfect testing ground for quantum gravity. 
Among these objects, BHs are of special interest due to the extreme curvature regimes they exhibit, as BHs represent regions of spacetime where the gravitational field becomes so intense that no signal or information, including light, can escape, rendering them inaccessible to direct observation. The event horizon delineates the ultimate limit beyond which escape is impossible. In classical BH solutions, a curvature singularity is at the core and is shielded by the event horizon; however, Several theoretical constructions propose regular BH models in which the central curvature singularity is resolved or avoided, while the event horizon itself remains classically regular. The temporal evolution of quantum entanglement in eternal BHs embedded in anti de Sitter spacetime, along with the dependence of entropy on interior geometric properties, was analyzed in~\cite{hartman2013time}. In addition, Engelhardt and Wall \cite{engelhardt2015quantum} proposed quantum extremal surfaces, which provide a method for calculating holographic entanglement entropy that accounts for quantum corrections beyond the traditional GTR description. A variety of BH solutions incorporating asymptotic safety effects were explored in previous studies \cite{cai2010black,zhang2018corrected,pawlowski2024effective}. 
From a quantum point of view, BHs are not entirely black. Hawking proved that quantum fluctuations near the event horizon produce thermal radiation \cite{hawking1978quantum}, which causes a slow loss of mass and a rise in temperature. This has significant effects on the thermodynamic behaviour of small black holes \cite{gour2003thermal}. The  related thermodynamic corrections and quantum effects have also been investigated for higher-dimensional, charged, and rotating BH configurations ~\cite {pourhassan2017thermodynamics,javed2023thermal,yousaf2024fuzzy}. 

A major observational milestone was achieved with the first image of the supermassive BH M87* in 2019 \cite{akiyama2019first,akiyama2019firstb,akiyama2019firsta}, followed by the imaging of the Milky Way’s central BH, $\operatorname{Sgr A}^*$, in $2022$ \cite{akiyama2022first}. These EHT observations provided direct visual confirmation of the BHs by detecting a dark central region, known as the BH shadow. This shadow corresponds to the region bounded by unstable photon orbits and is closely associated with the critical photon sphere \cite{guo2021universal}. The morphology and size of the shadow constitute a distinctive signature of radiation emitted by accretion flows surrounding supermassive compact objects \cite{johannsen2013photon} and are highly sensitive to the underlying spacetime geometry. Consequently,  investigations of BH shadows have been carried out within a wide range of gravitational models, significantly extending earlier foundational studies \cite{synge1966escape,johannsen2010testing,abdujabbarov2016shadow,zare2024shadows}.

Quasi-periodic oscillations (QPOs) observed in the X-ray emission spectra of microquasars offer deep insight into the physical nature of compact binary systems \cite{pasham2019loud,smith2021confrontation,singh2022low}. Pasham \emph{et al.} \cite{pasham2025using} expanded these studies by using infrared dust echo techniques to systematically detect bright quasi-periodic eruption sources. Evidence for a binary BH configuration inferred from QPO has been discussed \cite{pasham2024case}. Intense X-rays are released as accreting matter spirals inward toward the innermost stable circular orbit (ISCO) due to increased viscous dissipation and frictional factors \cite{wilkins2011determination}. In this regard, QPOs are essential for constraining intrinsic BH properties such as mass, spin, and electric charge, as well as for examining the microphysics of accretion fluxes\cite{DONMEZ2026170350}. 

The study of such systems relies on complementary observational techniques: spectroscopic analyses, which focus on the energy and frequency distribution of emitted photons, and timing analyses, which examine temporal variations in photon flux. Together, these methods form a robust framework for investigating accretion dynamics and strong-field gravity effects \cite{remillard2006x}. 
In a related theoretical development, Ashraf \emph{et al.} \cite{ashraf2025thermal} examined the dynamics of test particles orbiting a Schwarzschild BH embedded in a Dehnen-type dark matter halo, demonstrating how variations in the BH mass, halo radius, and central density influence particle motion. Closed-form expressions were derived for the particle energy, angular momentum, and effective potential. The authors further explored epicyclic oscillations near the equatorial plane by analyzing radial, vertical, and orbital frequencies, as well as periastron precession. Additionally, high-energy particle collisions near the event horizon were investigated through the evaluation of centre-of-mass energy, revealing a pronounced sensitivity of particle dynamics to the underlying model parameters. Complementary analyses of orbital motion and QPO diagnostics around rotating hairy BHs in Horndeski gravity were presented in~\cite{ashraf2025orbital}. Overall, extensive research has been devoted to understanding QPO phenomena in the strong-gravity regime surrounding BHs, underscoring their significance as powerful observational probes of relativistic astrophysics \cite{rezzolla2004new,belloni2012high}.

Low-frequency QPOs (LFQPOs) observed in a wide class of astrophysical systems cover a broad frequency range, extending from a few millihertz up to several tens of hertz. An interpretation of these signals requires an understanding of the complex accretion processes occurring near BHs, as such dynamics encode BH properties, most notably their mass and spin. In Ref.~\cite{stuchlik2013multi}, a multi-resonance orbital model was developed for high-frequency quasi-periodic oscillations (HFQPOs), demonstrating its potential to constrain the spin parameters of both BHs and neutron stars with high precision. 
The quasi-harmonic oscillatory behaviour of charged particles orbiting a Schwarzschild BH embedded in a uniform magnetic field was analysed in \cite{kolovs2015quasi}, while a comprehensive investigation of particle dynamics and associated QPO characteristics around an Euler-Heisenberg BH surrounded by a cold dark matter halo was presented in \cite{mustafa2025particle}. To account for the origin of QPOs, several theoretical mechanisms have been proposed, including diskoseismic modes, localized hot-spot models, nonlinear resonance phenomena, and warped accretion disk configurations \cite{rezzolla2013relativistic}. 

More recently, the geometric structure, particle motion, and thermodynamic behaviour of a non-rotating Frolov BH \cite{bouzenada2025barrow}. Using the Hamiltonian formalism, the authors assessed the influence of the electric charge and additional model parameters on orbital stability, the effective potential, and test-particle dynamics. Their analysis revealed that increasing the charge $Q$ and the parameter $\alpha$ drives the innermost stable circular orbits (ISCOs) closer to the event horizon, strengthens the effective gravitational attraction, and diminishes orbital stability. From a thermodynamic standpoint, the BH configuration exhibited positive temperature and entropy, while the behaviour of the specific heat signalled the presence of phase transitions governed by $Q$, $\alpha$, and the Barrow entropy parameter.

Observational studies, particularly those based on the source XTE~J1550$-$564, have led to the classification of low-frequency quasi-periodic oscillations (LFQPOs) into three distinct categories \cite{homan2001correlated,belloni2024fast}. Among these, Type A QPOs are the least frequently detected and typically emerge during transitional phases between hard and soft spectral states, having been reported in only a limited number of systems. Type B QPOs occur less frequently than Type C and exhibit a strong association with jet activity in certain BH binary systems. In contrast, Type C QPOs are the most prevalent class and are predominantly identified by detailed X-ray spectral and timing analyses \cite{motta2012discovery,motta2016quasi}.

Using Insight HXMT observations of the BH candidate GRS~1915$+$105, Liu \emph{et al.} \cite{liu2021testing} examined the evolution of the characteristics of LFQPO, focusing on the relationship between QPO frequency, inner radius of the accretion disk, and mass accretion rate. Their results revealed a pronounced positive correlation between the QPO frequency and the accretion rate, in agreement with and extending earlier AstroSat findings. Interpreting the observed QPO frequency as a relativistic dynamical frequency associated with a truncated accretion disk, the authors provided strong evidence supporting a rapidly spinning BH in GRS~1915$+$105 and emphasized the source’s relevance for future tests of general relativity. The dominance of Type C QPO behaviour in GRS~1915$+$105 has therefore made it a prime target for extensive observational and theoretical investigations aimed at uncovering the fundamental properties of BH environments and their associated accretion dynamics \cite{belloni2013discovery,misra2020identification,chauhan2024spectral}.

Building upon this body of work, the present study focuses on rotating black hole geometries described by an infrared-effective metric arising from the renormalization-group flow of asymptotically safe gravity, characterized by a scale-dependent gravitational coupling. Earlier investigations of quantum-modified Kerr spacetimes were carried out in \cite{reuter2011quantum}, where the classical Newtonian constant was replaced by a radially varying function induced by quantum gravitational effects. That analysis examined modifications to the horizon structure, ergoregions, static limit surfaces, and energy-extraction processes. In contrast, the present study employs an infrared-inspired functional form of the gravitational coupling within the asymptotic safety paradigm, enabling a more concrete and systematic analysis. This gravitational background allows the derivation of explicit analytic expressions for equatorial geodesics and facilitates a reexamination of the Penrose energy-extraction mechanism in the presence of quantum-gravitational corrections. Furthermore, we investigate accretion disk behavior and the emergence of QPOs around Kerr BHs in the infrared regime of asymptotically safe gravity.
Our relativistic hydrodynamic analysis of BHL accretion demonstrates that quantum effects play a crucial role in reshaping the shock cone structure formed in the vicinity of the BH, whereas frame-dragging associated with BH rotation governs the asymmetry of the cone, the quantum correction parameter effectively weakens the gravitational potential in this relativistic Kerr-like spacetime, leading to an increased opening angle, reduced post-shock compression, and a lower matter density within the shocked region. The temporal evolution of the mass accretion rate exhibits characteristic oscillatory modes confined to the shock cone; however, PSD analyses reveal that these modes give rise to LFQPOs, with their strength, coherence, and harmonic content sensitive to both the BH spin and the magnitude of the quantum corrections. Notably, intermediate values of rotation and quantum modification favour the development of coherent oscillations with nearly commensurate frequency ratios, whereas stronger quantum effects damp the oscillations and smear out well-defined spectral features. 

Throughout this work, we adopt geometric units $G=c=1$ and employ a spacetime signature $(-,+,+,+)$. Greek indices are taken to run from $0$ to $3$, while physical constants are explicitly retained whenever astrophysical interpretations are involved.\\

The structure of our manuscript is as follows: 
In Sec .~\ref {Kerrmetric_safegravity}, we introduce the Kerr metric in the infrared limit of asymptotically safe gravity, while Sec .~\ref {shock_structure} is devoted to the analysis of accretion dynamics and shock cone formation in the infrared-modified Kerr spacetime. In Sec .~\ref {QPO_analysis}, we examine the impact of quantum gravitational corrections on QPOs through a detailed PSD analysis. Furthermore, Sec.~\ref {compareWith_Obs} focuses on the identification of observationally relevant QPO frequencies extracted from the numerically obtained PSDs.

%%%%%%%%%%%%%%%%%%%%%%%%%%%%%%%%%%%%%%%%%%%%%%%%%%%%%%%%%%%%%%%%%%%%%%%%%%%
\section{Kerr metric in the infra-red limit of asymptotically safe gravity theory}\label{Kerrmetric_safegravity}
%%%%%%%%%%%%%%%%%%%%%%%%%%%%%%%%%%%%%%%%%%%%%%%%%%%%%%%%%%%%%%%%%%%%%%
The line element describing the geometry of a rotating and embedded BH is given by \cite{haroon2018effects}

\begin{eqnarray}\label{maineq}
d s^2=  -\left(1-\frac{2 M r}{\Sigma}\left(1-\frac{\xi}{r^2}\right)\right) d t^2 \;\;\;\;\;\;\;\;\;\;\;\;\;\;\;\;\;\;\;\;\; \nonumber\\
 -\frac{4 a M r \sin ^2 \theta}{\Sigma}\left(1-\frac{\xi}{r^2}\right) d t d \phi+\frac{\Sigma}{\Delta} d r^2+\Sigma d \theta^2 \nonumber\\
 +\sin ^2 \theta\left[r^2+a^2+\frac{2 a^2 M r}{\Sigma} \sin ^2 \theta\left(1-\frac{\xi}{r^2}\right)\right] d \phi^2,  
\end{eqnarray}
where $\Sigma=r^2+a^2 \cos ^2 \theta$ and $\Delta=r^2-2 M r+\frac{2 M \xi}{r}+a^2$. This metric reduces to its static, spherically symmetric form when $a \rightarrow 0$. To aid the reader's understanding and provide stronger grounds for the results computed in the rest of the sections, a detailed derivation of the metric in Eq. (\ref{maineq}) using the technique in \cite{azreg2014generating} is presented in the appendix.

\begin{figure*}
  \center
   \psfig{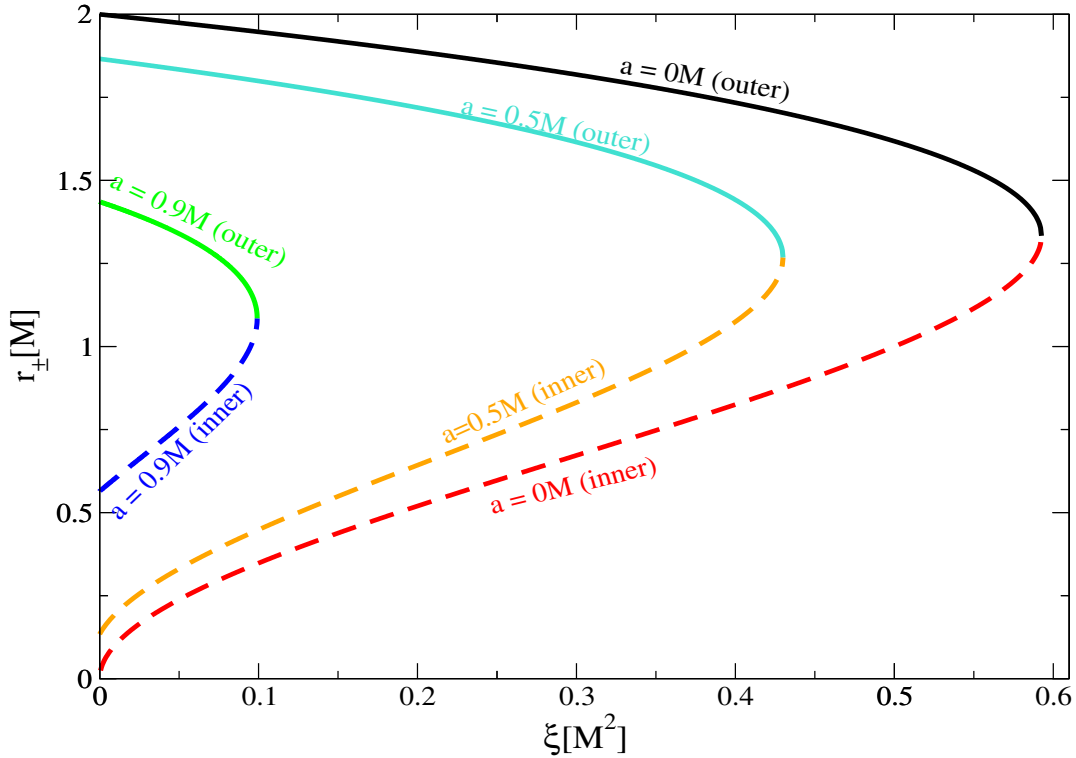}   
\caption{ Variation of the outer ($r_{+}$, solid lines) and inner ($r_{-}$, dashed lines) black hole horizons as a function of the quantum correction parameter $\xi$ for the Kerr metric in the infra-red limit of asymptotically safe gravity. The horizons are shown separately for each black hole spin parameter $a = 0$, $0.5$, and $0.9\,M$, as listed in Table \ref{Initial_data}. An increase in $\xi$ causes the inner and outer horizons to approach each other and eventually merge at a critical value, beyond which no horizon exists.
}
\label{horizon}
\end{figure*}

For the Kerr metric in the infrared limit of asymptotically safe gravity, the locations of the black hole horizons are determined by the roots of
$\Delta(r)=0,$ 
which correspond to null hypersurfaces where
$g^{rr}=\frac{\Delta}{\Sigma}=0.$
Hence, the horizon radii follow directly from the condition $\Delta(r)=0$. Fixing the black hole mass to $M=1$, we solve the horizon equation numerically for different values of the spin parameter $a$, obtaining the inner ($r_{-}$) and outer ($r_{+}$) horizons as functions of the quantum correction parameter $\xi$. The resulting behaviour of the horizons is shown in Fig.~\ref{horizon}, where the horizon radii are plotted as functions of $\xi$ for $a=0$, $0.5$, and $0.9$. For each value of $a$, increasing $\xi$ causes the inner and outer horizons to approach each other, and this convergence persists up to a critical value $\xi_{\mathrm{crit}}$. At this point, the two horizons coincide, forming a degenerate (extremal) horizon. For $\xi>\xi_{\mathrm{crit}}$, the horizon equation admits no positive real roots, indicating the absence of an event horizon and the emergence of a horizonless spacetime, corresponding to a naked singularity.

\begin{figure*}[h!]
  \vspace{1cm}
  \center
   \psfig{file=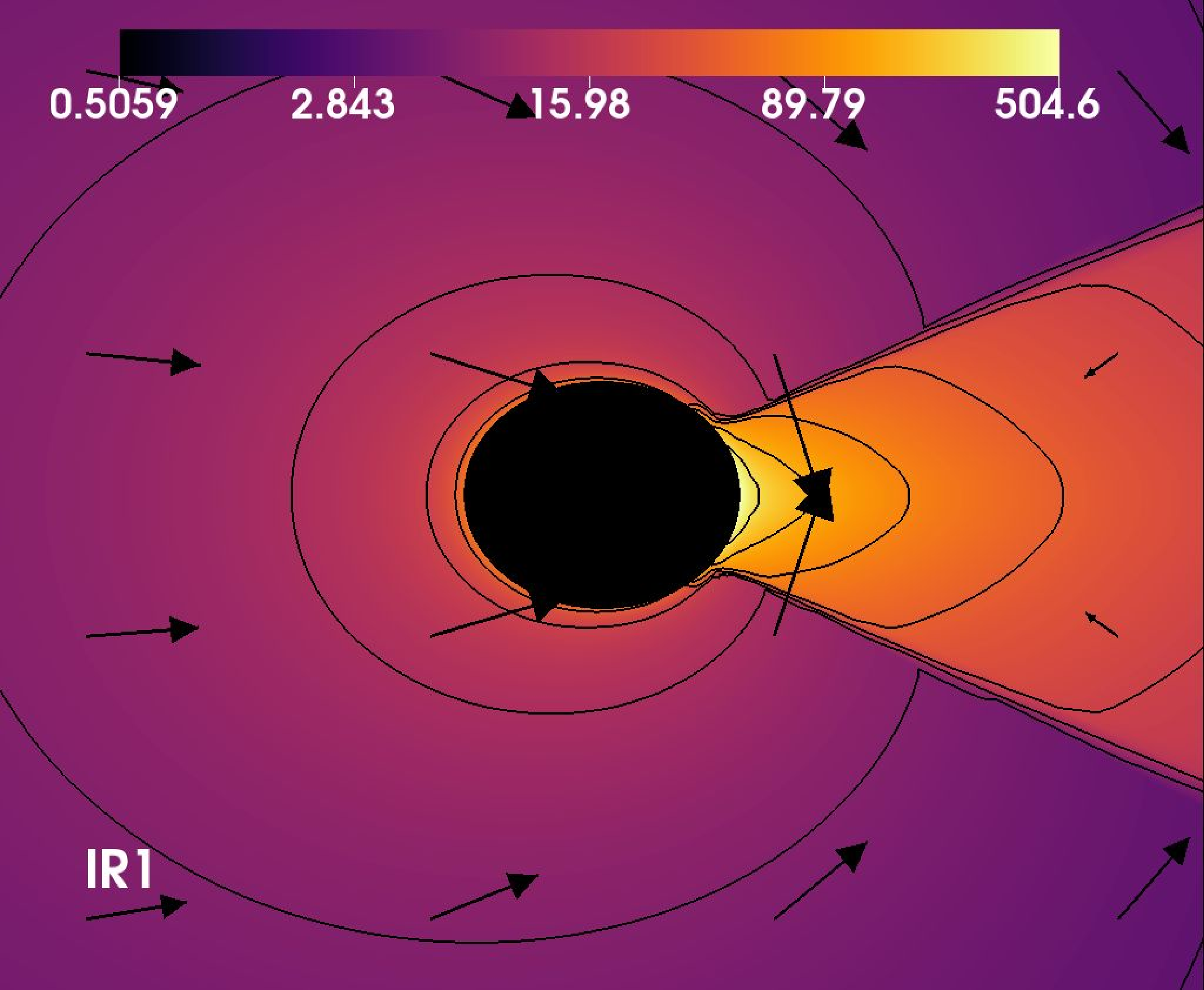,width=7.5cm,height=7.0cm} 
   \psfig{file=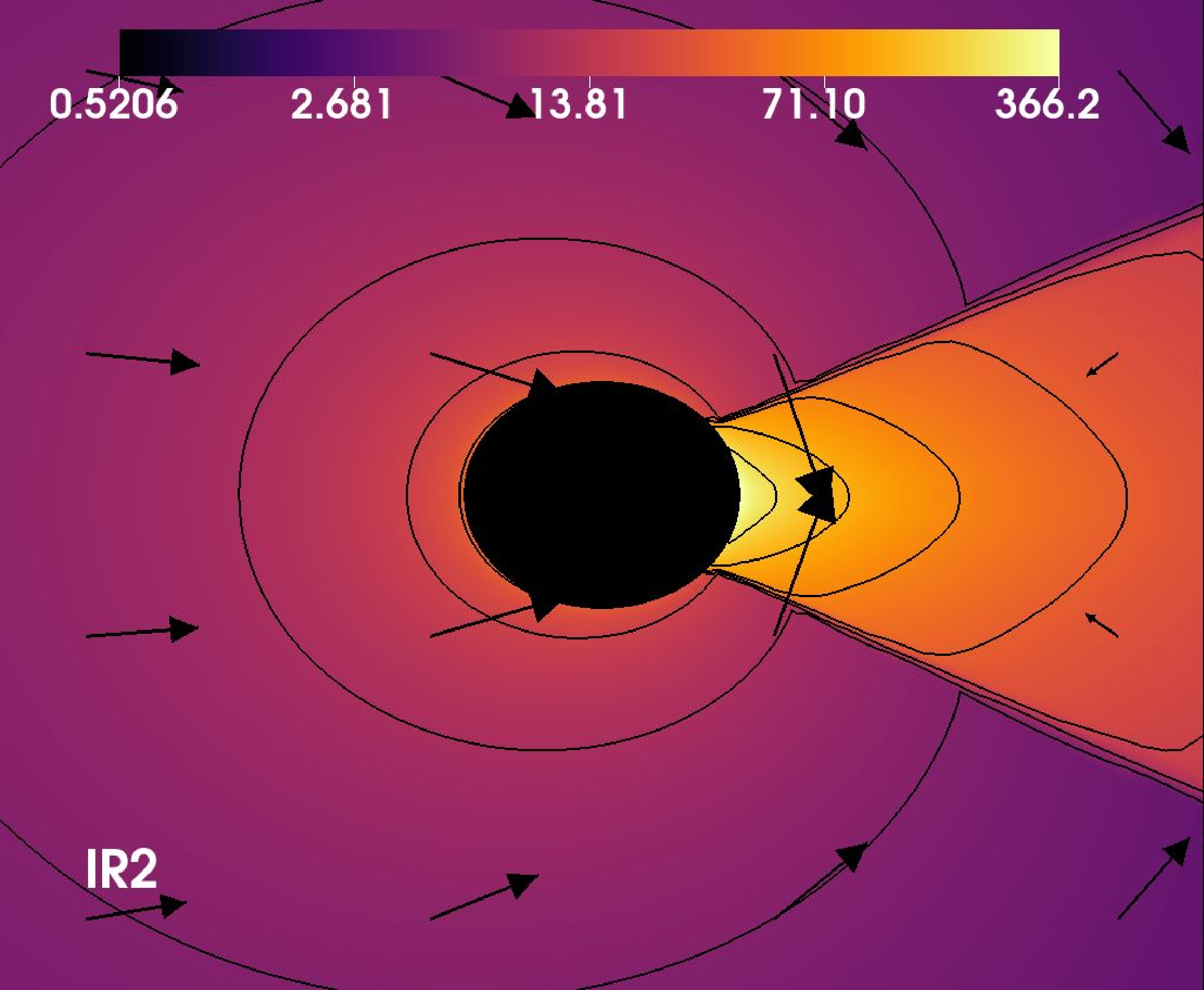,width=7.5cm,height=7.0cm} \\  
   \psfig{file=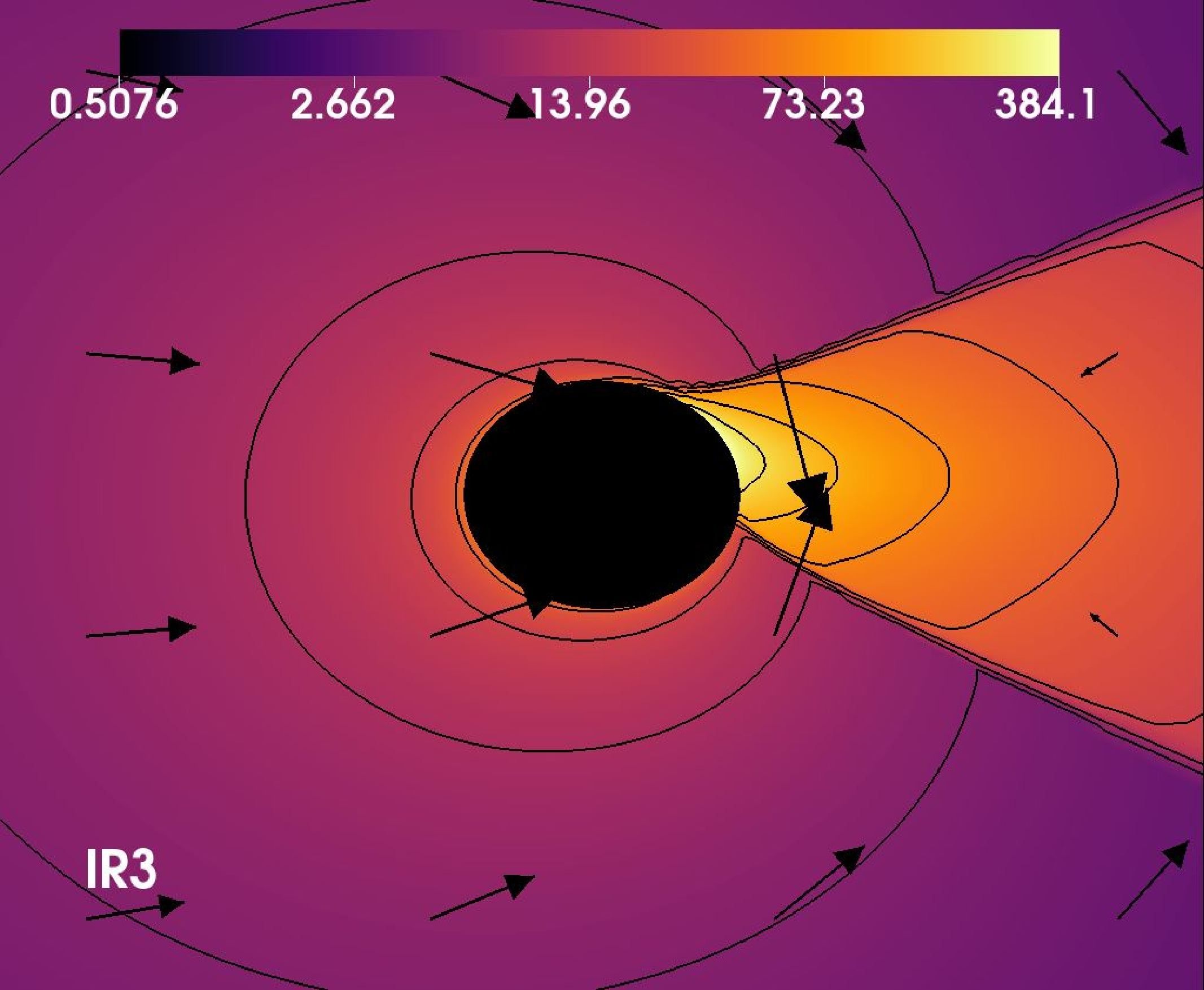,width=7.5cm,height=7.0cm} 
   \psfig{file=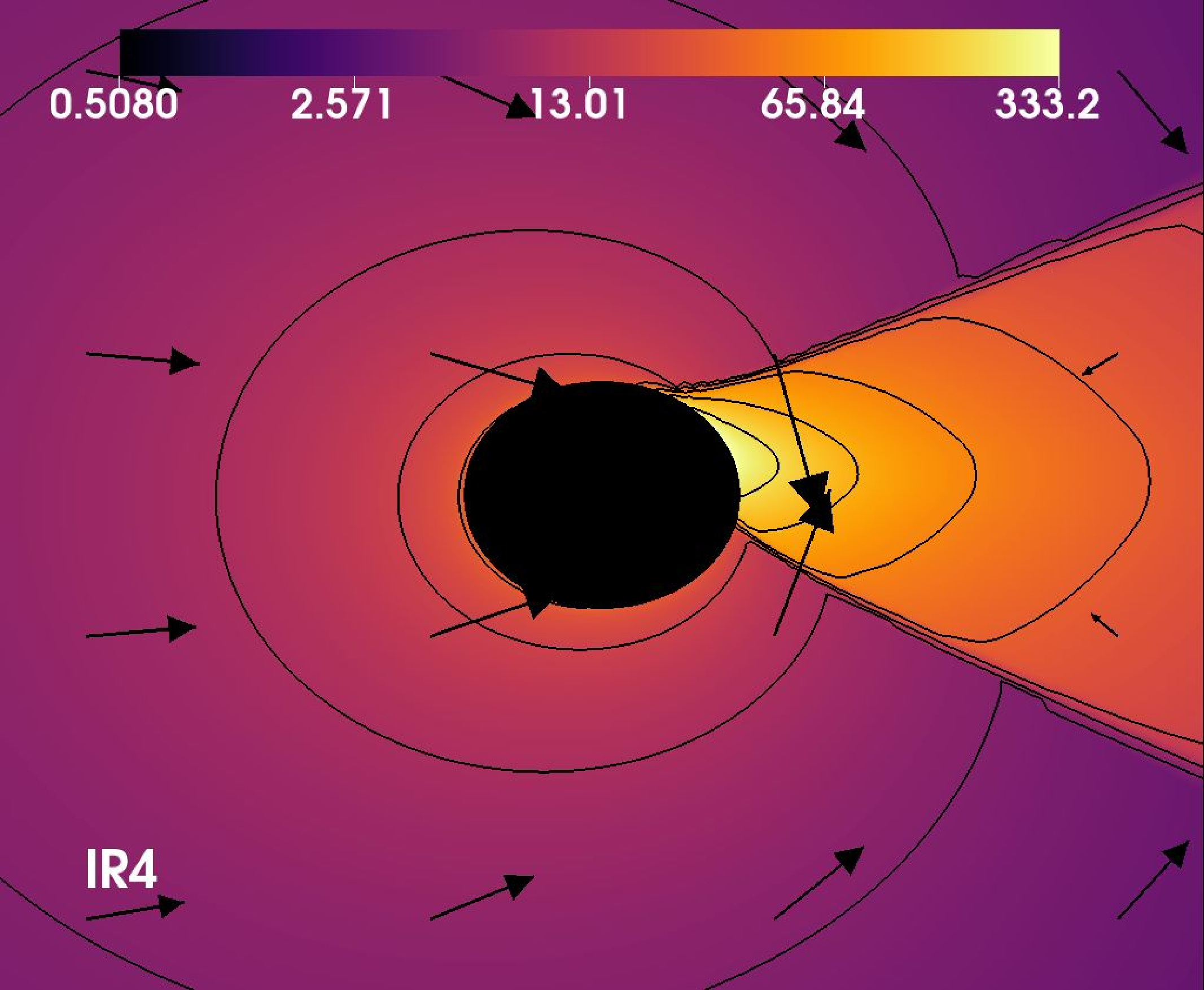,width=7.5cm,height=7.0cm} \\  
   \psfig{file=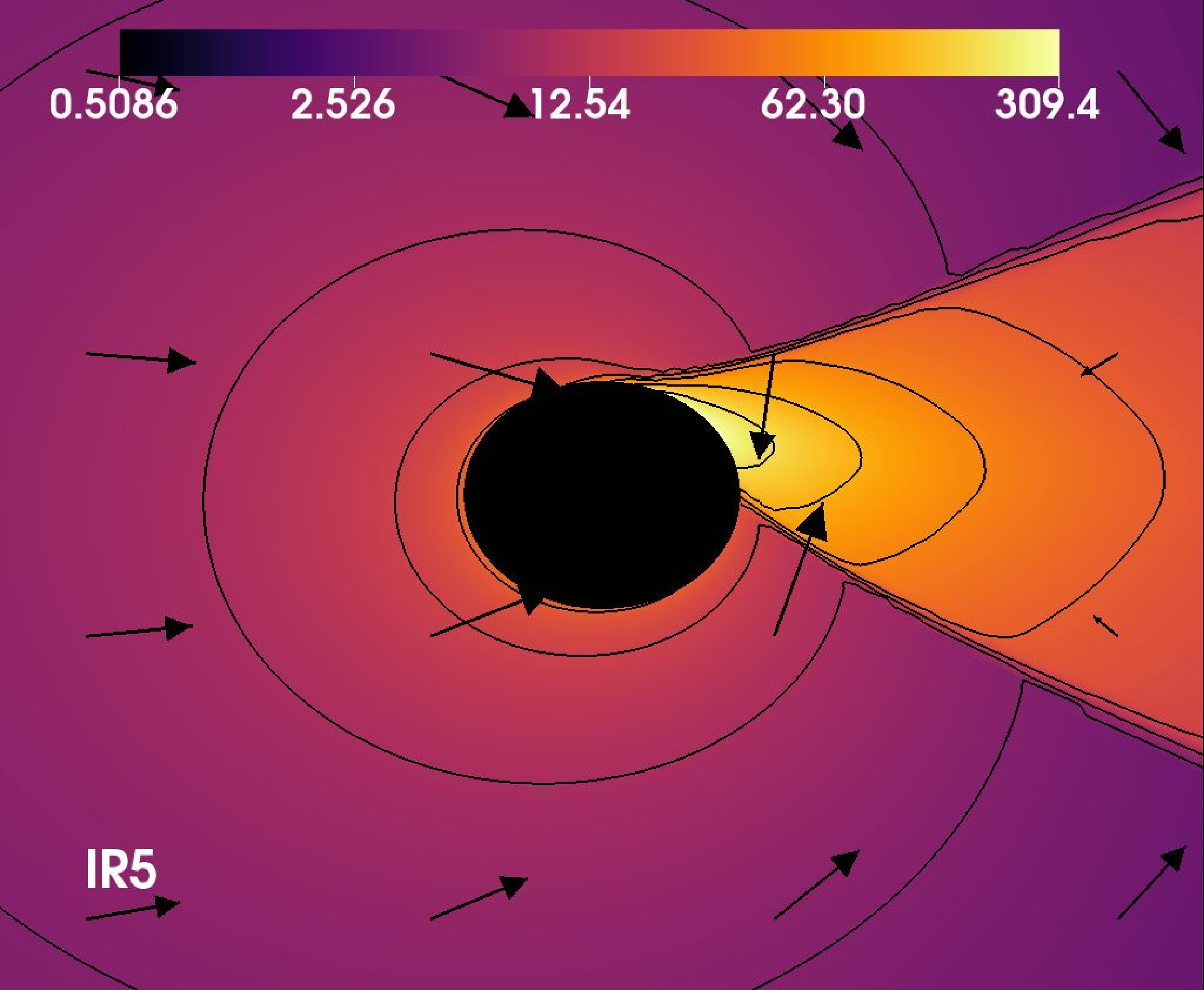,width=7.5cm,height=7.0cm} 
   \psfig{file=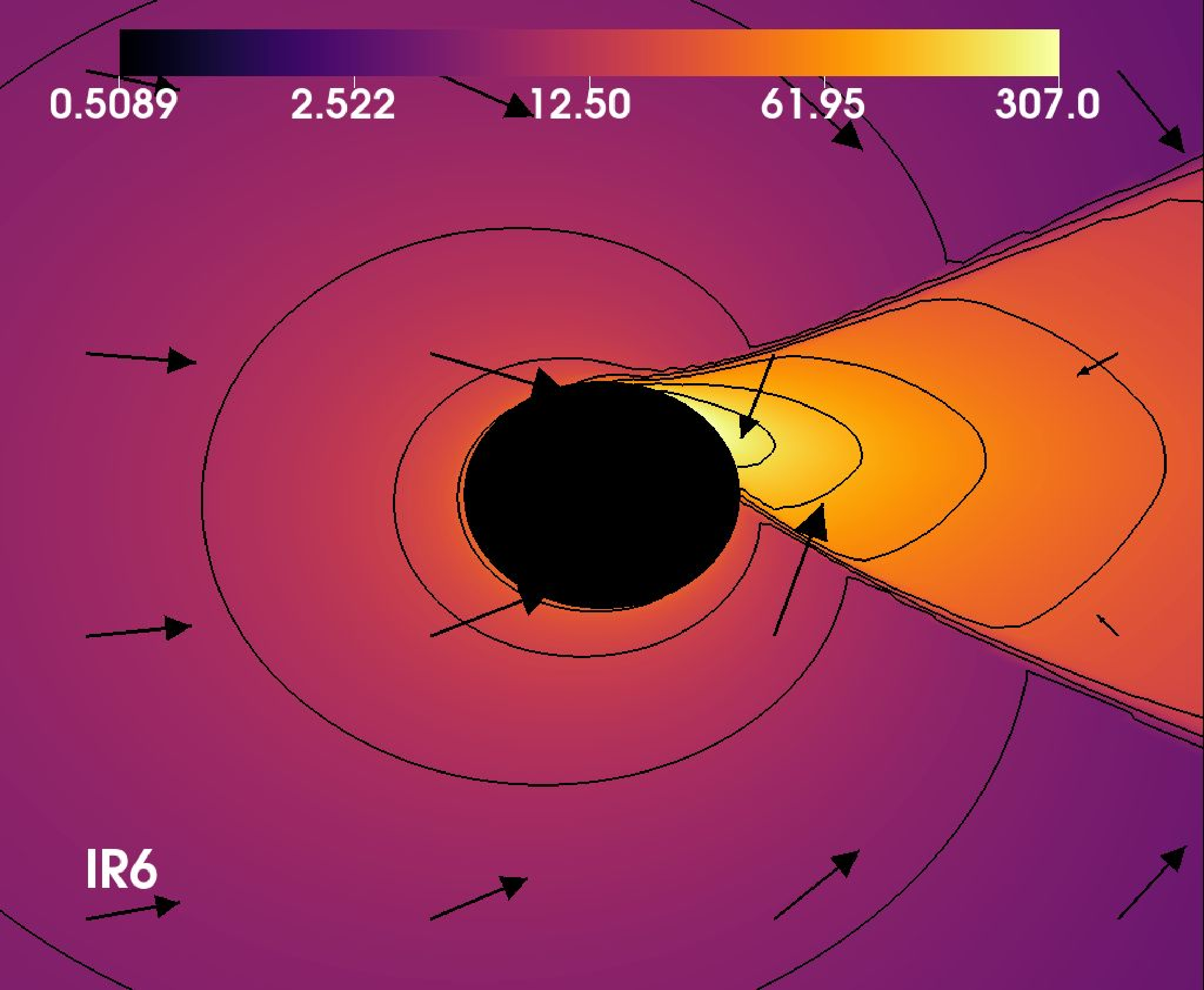,width=7.5cm,height=7.0cm}      
\caption{Color and contour plots of the shock cone and plasma structure formed via the BHL mechanism around a Kerr black hole in the infrared limit of asymptotically safe gravity are presented. Each panel corresponds to a different model listed in Table \ref{Initial_data}. The variation of the shock cone and its morphology in the strong gravitational field is shown for different values of the black hole spin parameter $a$ and the quantum correction parameter $\xi$. In addition, the velocity field is illustrated by vector plots, revealing the inflow of matter toward the black hole and the flow structure within and around the shock cone.}
\vspace{1cm}
\label{color_plots}
\end{figure*}

\begin{figure*}
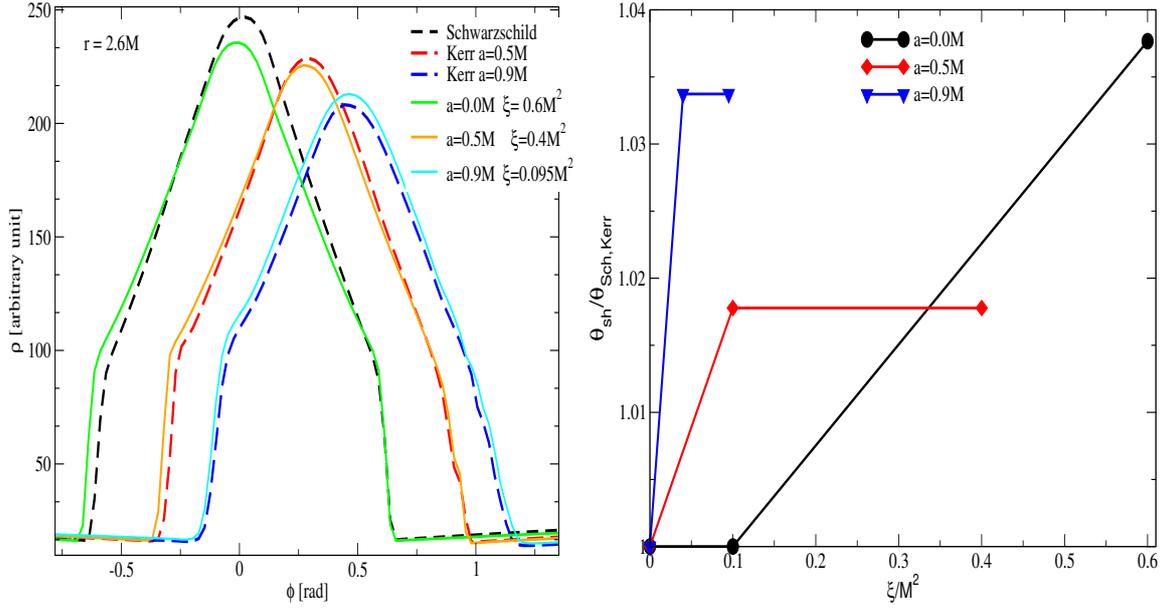

  \center
   \psfig{file=den_phi_r26.eps,width=7.5cm,height=8.0cm} 
   \psfig{file=shock_angle_r26.eps,width=7.5cm,height=8.0cm}    
\caption{ Left panel: The azimuthal variation of the density of the shock cone formed around the black hole, computed at $r = 2.66M$, i.e., very close to the black hole horizon. The profiles are shown for different values of the spin parameter $a$ and the quantum correction parameter $\xi$, and are also plotted for the corresponding Schwarzschild and Kerr models with the same spin values for comparison. Right panel: Variation of the normalized shock cone opening angle as a function of $\xi$ for different spin configurations. The shock opening angle is normalized with respect to the corresponding Schwarzschild or Kerr case for each spin value, thereby highlighting deviations from the classical black hole models.
}
\label{density_angle}
\end{figure*}

\begin{figure*}[h!]
  \center
  \psfig{file=acc_rate_r23.eps,width=15.0cm,height=10.0cm}  \\
   \psfig{file=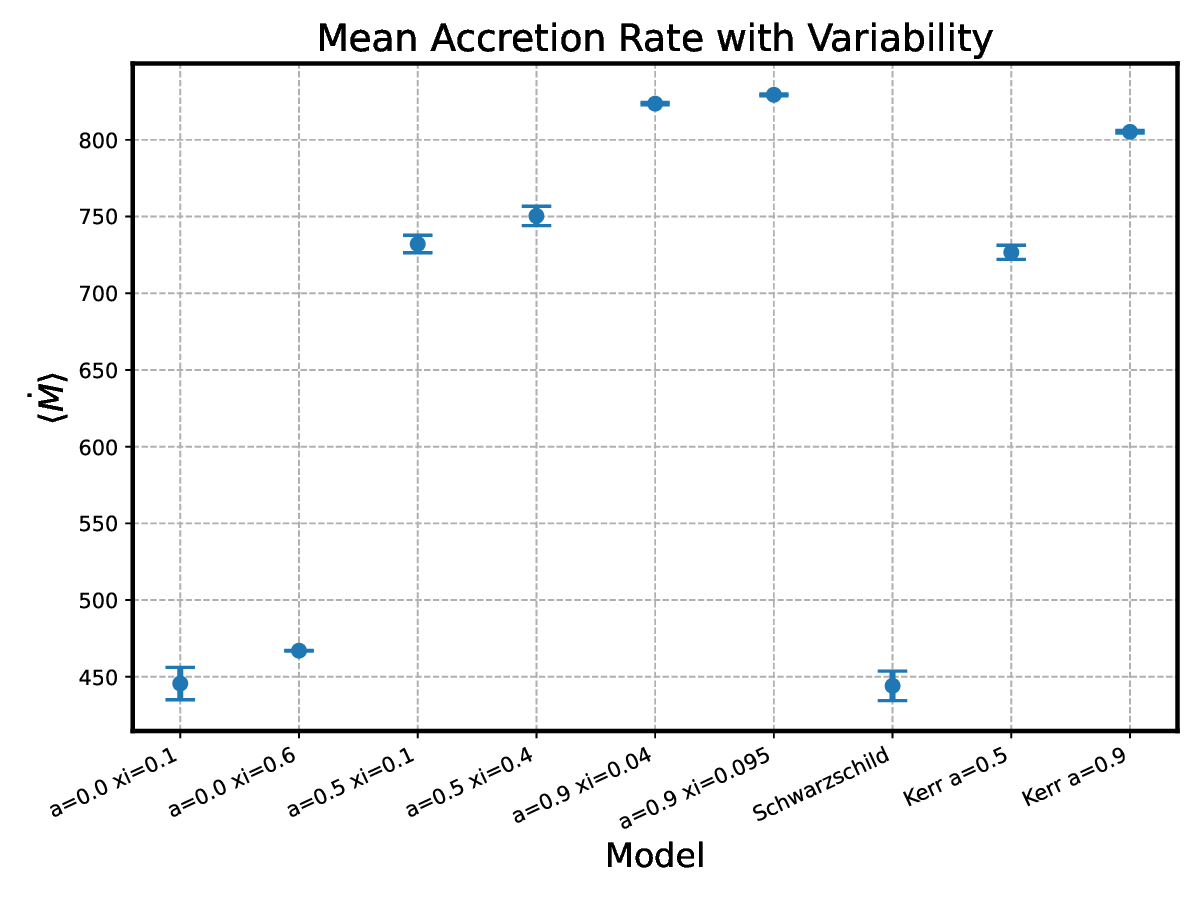,width=7.5cm,height=8.0cm} 
   \psfig{file=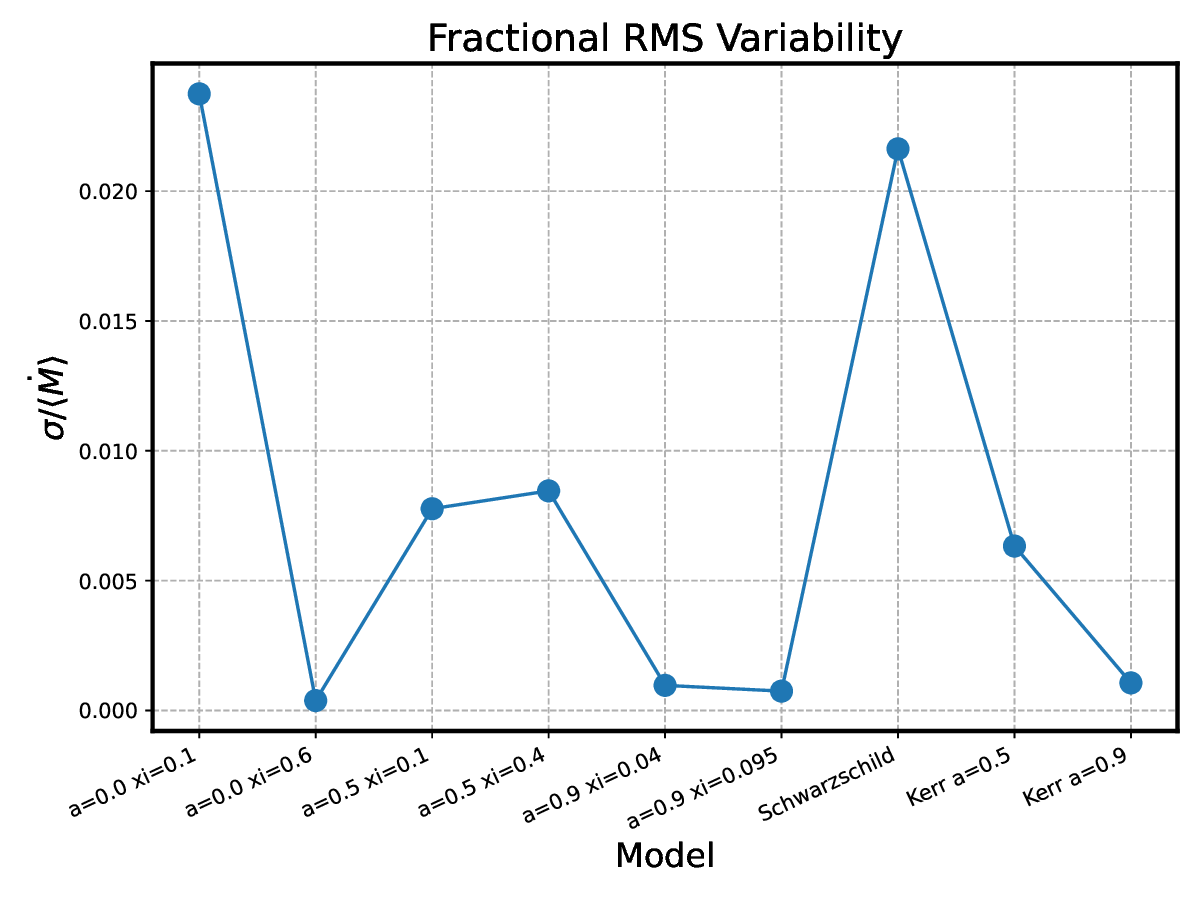,width=7.5cm,height=8.0cm}    
\caption{Numerical results for the mass accretion properties of the quantum-corrected Kerr black hole in the infra-red limit of asymptotically safe gravity. Top panel: The time evolution of the mass accretion rate $dM/dt$ computed at the location closest to the black hole horizon, $r = 2.3M$, is shown for different values of the spin parameter $a$ and the quantum correction parameter $\xi$, together with the Schwarzschild and Kerr cases, after the shock cone reaches a quasi-steady state. Bottom left panel: The mean mass accretion rate $\langle \dot{M} \rangle$ calculated from the corresponding accretion rate shown in the top panel. The error bars represent the standard deviation of temporal fluctuations. Bottom right panel: The model-dependent variation of the fractional RMS variability, which quantifies the relative strength of fluctuations in the accretion rate for each model.
}
\label{acc_r23}
\end{figure*}

\begin{figure*}[h!]
  \vspace{1cm}
  \center
  \psfig{file=acc_rate_r611.eps,width=15.0cm,height=10.0cm}  \\
   \psfig{file=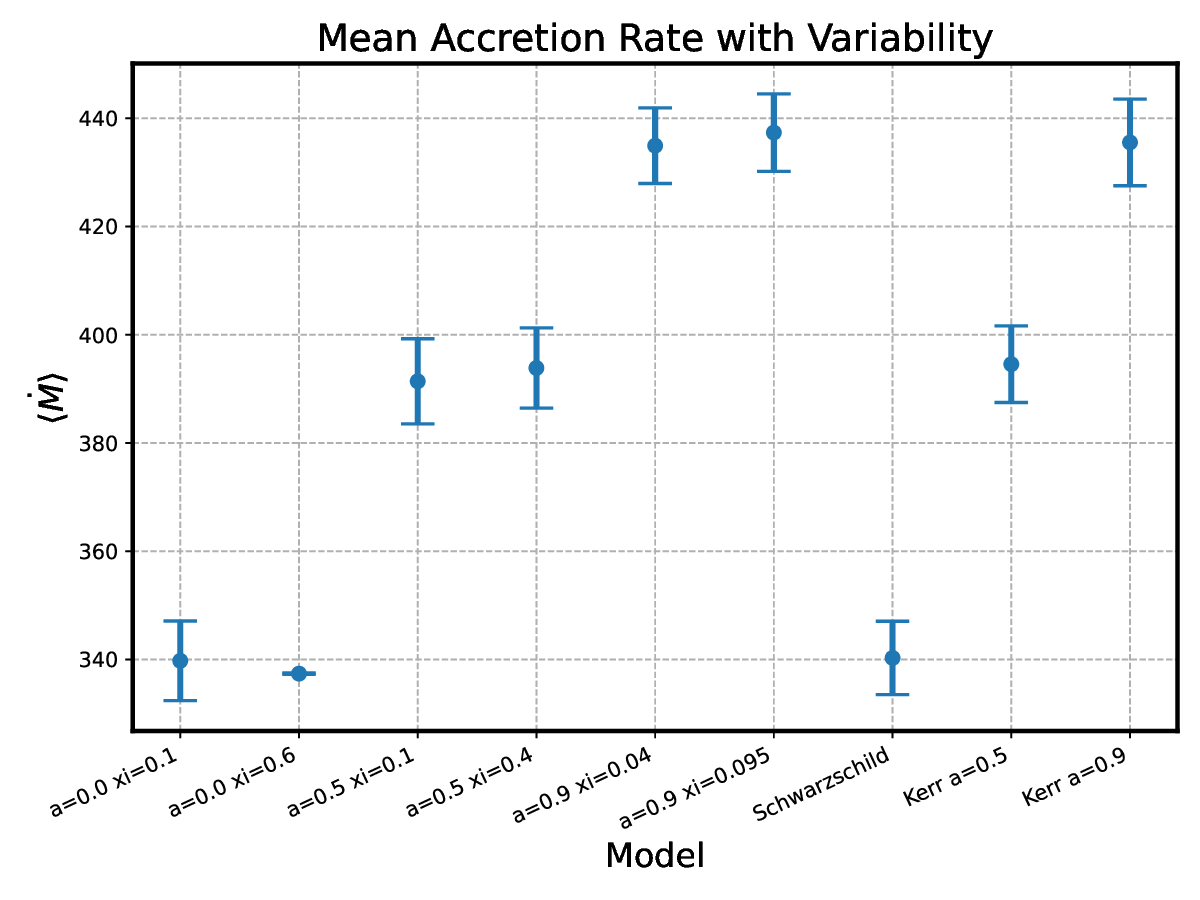,width=7.5cm,height=8.0cm} 
   \psfig{file=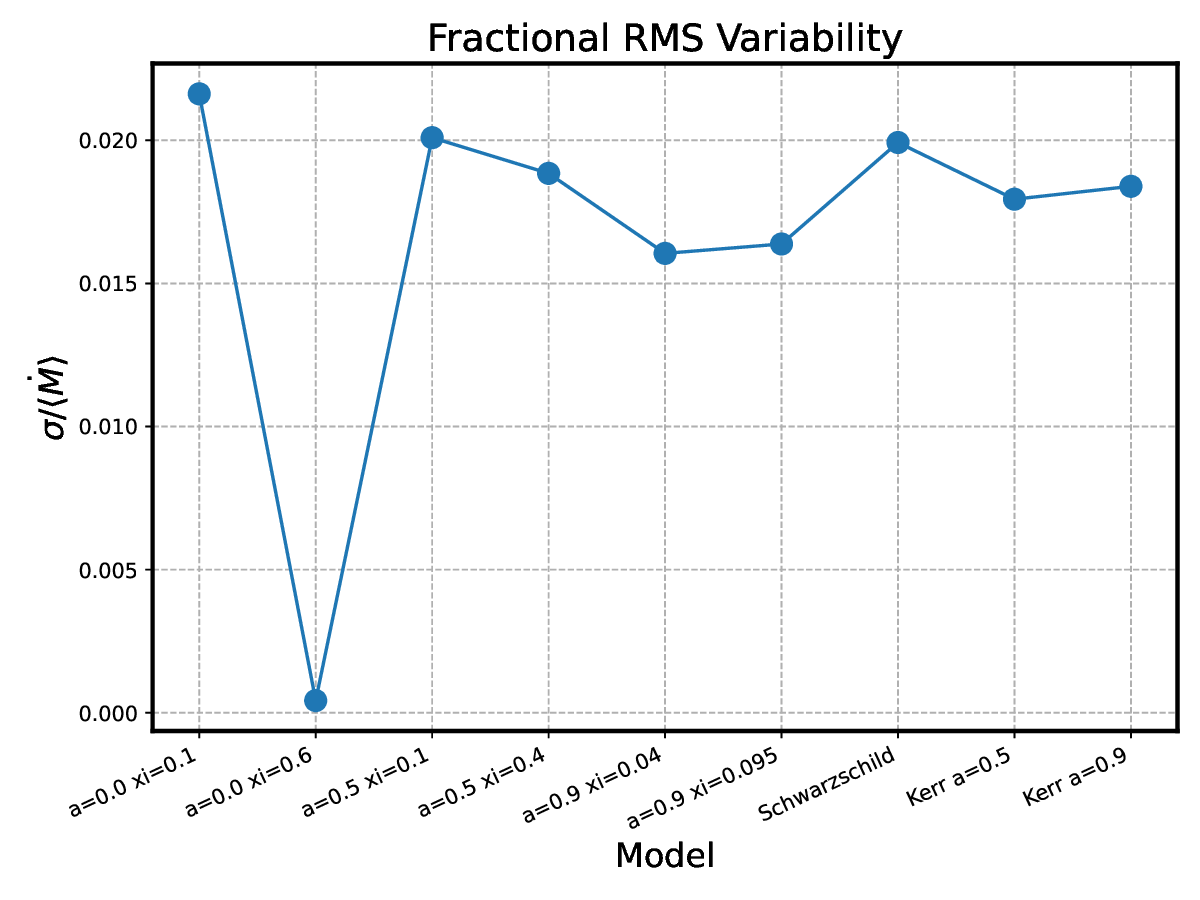,width=7.5cm,height=8.0cm}    
\caption{Numerical results for the mass accretion properties of the quantum-corrected Kerr black hole in the infra-red limit of asymptotically safe gravity. The figure is the same as Fig.\ref{acc_r23}, but in this case the mass accretion rate $dM/dt$, the mean accretion rate $\langle \dot{M} \rangle$, and the fractional RMS variability are computed at $r = 6.11M$, corresponding to a region farther from the immediate vicinity of the black hole.
}
\vspace{1cm}
\label{acc_r611}
\end{figure*}

\begin{figure*}
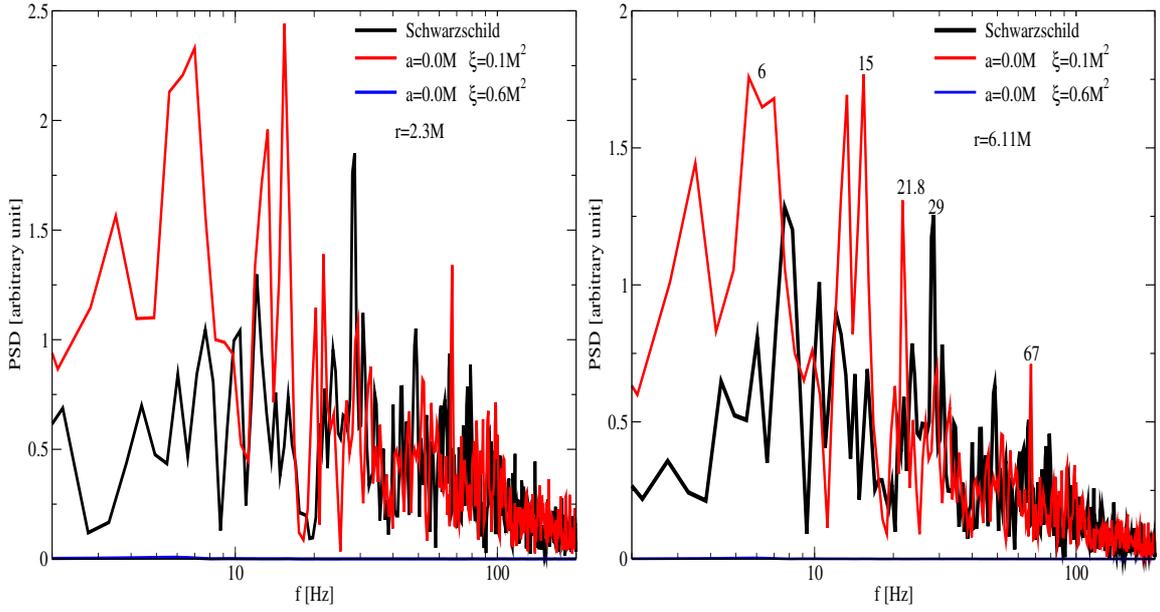

  \center
   \psfig{file=PSD_ALL_r23a00.eps,width=7.5cm,height=8.0cm} 
   \psfig{file=PSD_ALL_r611a00.eps,width=7.5cm,height=8.0cm}    
\caption{ The PSD analysis is performed for both the Schwarzschild black hole and quantum-corrected black holes with different values of the parameter $\xi$, using the temporal oscillations of the mass accretion rate after it reaches saturation around a nonrotating black hole. The left panel shows the PSD computed from the mass accretion rate extracted at $r=2.3M$, corresponding to the region where matter falls toward the black hole, while the right panel presents the PSD evaluated at $r=6.11M$. The figures provide a comparative illustration of the variations in the QPO characteristics as a function of the black hole spin parameter $a$ and the quantum-correction parameter $\xi$. All frequencies are scaled to a black hole mass of $M = 10\,M_\odot$.
}
\label{QPOs_a00}
\end{figure*}

\begin{figure*}[h!]
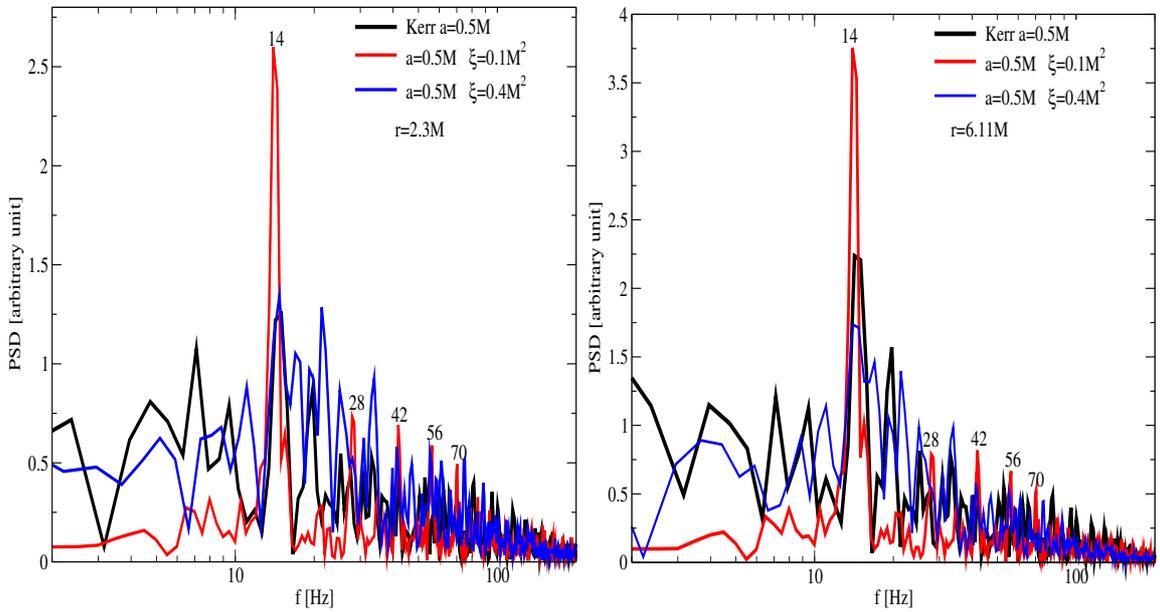

  \center
   \psfig{file=PSD_ALL_r23a05.eps,width=7.5cm,height=8.0cm} 
   \psfig{file=PSD_ALL_r611a05.eps,width=7.5cm,height=8.0cm}    
\caption{Same as Fig.\ref{QPOs_a00}, but in this case the PSD analysis is presented comparatively for a moderately rotating Kerr black hole with spin parameter $a=0.5M$ and the corresponding quantum-corrected black holes. 
}
\label{QPOs_a05}
\end{figure*}

\begin{figure*}[h!]
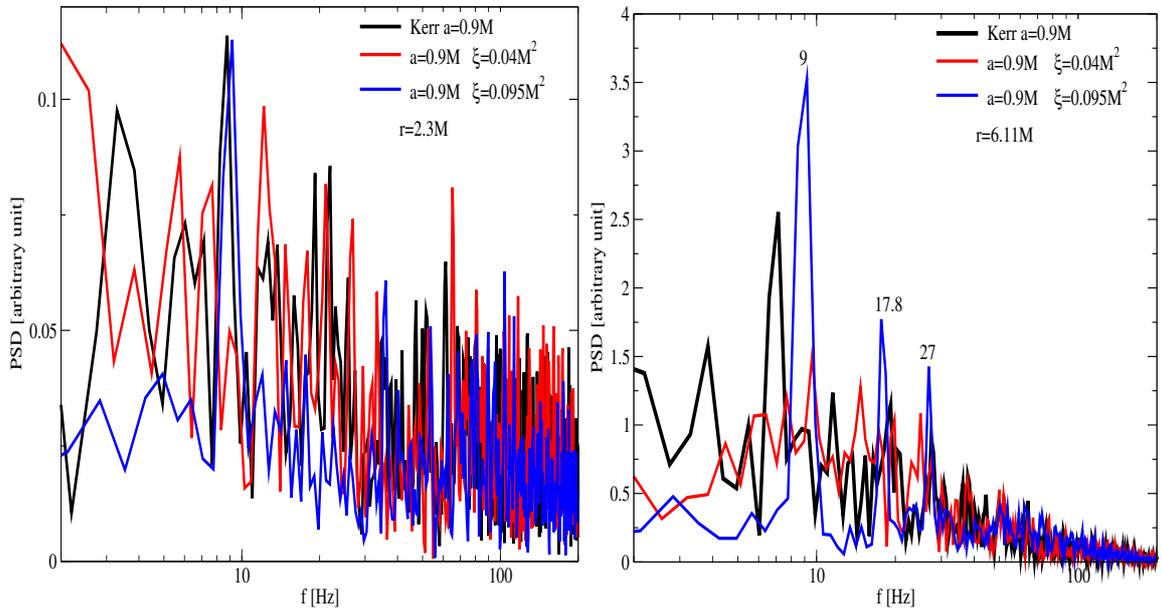

  \vspace{1cm}
  \center
   \psfig{file=PSD_ALL_r23a09.eps,width=7.5cm,height=8.0cm} 
   \psfig{file=PSD_ALL_r611a09.eps,width=7.5cm,height=8.0cm}    
\caption{Same as Figs.\ref{QPOs_a00} and \ref{QPOs_a05}, but in this case the PSD analyses are computed for rapidly rotating black holes with spin parameter $a=0.9M$.
}
\vspace{1cm}
\label{QPOs_a09}
\end{figure*}
%%%%%%%%%%%%%%%%%%%%%%%%%%%%%%%%%%%%%%%%%%%%%%%%%%%%%%%%%%%%%%%%%%%%%%
%%%%%%%%%%%%%%%%%%%%%%%%%%%%%%%%%%%%%%%%%%%%%%%%%%%%%%%%%%%%%%%%%%%%%%
\section{Accretion Dynamics and Shock Phenomenology in Infra-Red Modified Kerr Geometry}
\label{shock_structure}

In this section, we investigate the accretion dynamics forming around a black hole described by the Kerr metric in the infrared limit of asymptotically safe gravity. In order to achieve this, we numerically solve the general relativistic hydrodynamic (GRHD) equations with the spacetime metric given in Eq.\ref{maineq}. The evolution of the fluid is carried out using a high-resolution shock-capturing (HRSC) scheme, which enables an accurate and stable treatment of nonlinear relativistic flows as well as the formation of strong discontinuities such as shocks \cite{Donmez:2004ss,Donmez:2005mc}. The hydrodynamic equations are written in conservative form and are evolved on a fixed quantum-corrected Kerr spacetime background. As a result of these calculations, deviations from the classical Kerr solution induced by the quantum correction term are clearly revealed in the numerical results.

 In order to investigate the strong gravitational field regime, we employ the BHL accretion mechanism in this study, which allows matter to fall toward the black hole \cite{2012MNRAS.426.1533D,Donmez:2014bza,Mustafa:2025mkc}. At the outer boundary of the computational domain, the inflowing matter is initialised with asymptotically supersonic velocities directed toward the black hole, enabling the development of physically relevant accretion dynamics around the compact object. As matter approaches the black hole, the strong gravitational field focuses the flow, leading to the formation of a shock cone in the downstream region, that is, on the opposite side of the incoming hemisphere. The resulting cone is characterized by strong shock locations and enhanced density in the post-shock region. 

In relativistic settings, particularly in rotating black hole spacetimes, the structure and stability of the shock cone are influenced by frame-dragging effects and the detailed form of the gravitational attraction. Consequently, physical properties such as the shock cone's morphology, opening angle, and the density and velocity distributions of the matter confined within the cone play a crucial role in testing the Kerr black hole in the infrared limit of asymptotically safe gravity and in revealing its deviations from the classical Kerr solution.

The initial conditions for the numerical simulations assume matter with uniform density, pressure, and velocity, specified at large distances from the black hole. The flow is initialised in the equatorial plane by injecting matter from the upstream region toward the black hole along a fixed direction. In this way, the accretion flow can evolve naturally and form a quasi-steady-state structure around the black hole.

The parameters defining the spacetime geometry surrounding the specific black hole models considered in this study are summarised in Table \ref{Initial_data}. As depicted in Table \ref{Initial_data}, different values of the black hole spin parameter $a$ are explored together with the corresponding quantum correction parameters $\xi$. These parameters determine the horizon structure of the spacetime, yielding configurations with distinct inner and outer horizons, as well as horizonless solutions for sufficiently large values of $\xi$. To facilitate a direct comparison with general relativity, the results obtained for the Kerr metric in the infrared limit of asymptotically safe gravity are analysed alongside the corresponding Schwarzschild and classical Kerr solutions for each value of the spin parameter.

\begin{table}[h]
\centering
\caption{Inner and outer horizons for the Kerr black hole in the infra-red limit of asymptotically safe gravity for different values of the spin parameter $a$ and the quantum correction parameter $\xi$ with $M=1$.} \label{Initial_data}
\begin{tabular}{c c c c}
\hline
\hline
Model & $a\,(M)$ & $\xi\,(M^{2})$ & Horizons $(r/M)$ \\
\hline
IR1 & 0.0 & 0.10  & $r_{-}=0.34,\;\;\;\; r_{+}=1.95$ \\
IR2 & 0.0 & 0.6  & No horizon \\
IR3 & 0.5 & 0.10  & $r_{-}=0.70,\;\;\;\; r_{+}=1.16$ \\
IR4 & 0.5 & 0.40  & $r_{-}=1.08,\;\;\;\; r_{+}=1.44$ \\
IR5 & 0.9 & 0.04  & $r_{-}=0.72,\;\;\;\; r_{+}=1.36$ \\
IR6 & 0.9 & 0.095 & $r_{-}=1.00,\;\;\;\; r_{+}=1.17$ \\
\hline
\hline
\end{tabular}
\end{table}

In this section, we discuss the dependence of the shock cone morphology and accretion dynamics formed around a Kerr black hole in the infra-red limit of asymptotically safe gravity. At the same time, by comparing our results with those corresponding to the classical Schwarzschild and Kerr solutions, the physical impact of $\xi$ can be more clearly identified, and the parameter ranges in which the infra-red limit of asymptotically safe gravity may be observationally tested can be delineated. Fig.~\ref{color_plots} presents the formation of the shock cone and its morphology around the black hole. In addition, the plasma structure developing around the black hole, together with the shock cone, is shown as shaped by the BHL accretion mechanism. The colored maps in Fig.~\ref{color_plots}  represent the density distribution of the accreting matter, while the overlaid contour lines denote the corresponding iso-density surfaces. The velocity field, illustrated by arrows, shows both the variation in the magnitude of the flow velocity and the direction of matter motion in the vicinity of the black hole.

In the top row of Fig.~\ref{color_plots}, the effect of the quantum correction parameter $\xi$ is demonstrated for a non-rotating black hole ($a=0M$) by comparing shock cone structures obtained for different values of $\xi$. In the left panel, corresponding to the IR1 model, the spacetime admits both inner and outer horizons, and the accretion flow develops a well-defined and narrow shock cone in the downstream region. Due to strong collimation of matter, a dense post-shock region is formed, and the velocity vectors indicate a smooth gravitational focusing of the inflow toward the black hole. On the other hand, for the IR2 model with the same spin parameter but a larger value of $\xi$, we have a naked singularity spacetime . In this case, the resulting shock cone is broader and exhibits weaker collimation of matter. The density contrast along the shock is reduced, and the velocity field indicates a less efficient focusing of the inflowing material. It demonstrates that increasing the quantum correction parameter $\xi$ in non-rotating configurations effectively softens the gravitational attraction, thereby weakening shock compression and modifying the accretion flow.

The middle row of Fig.~\ref{color_plots}  shows the IR3 and IR4 models for a moderately rotating black hole with spin parameter $a=0.5M$, illustrating the combined influence of rotation and quantum corrections. As is well known, black hole rotation induces a pronounced azimuthal asymmetry of the downstream shock cone through frame-dragging effects. When comparing the IR3 and IR4 models, we find that for the larger value of $\xi$ (IR4), the shock opening angle widens and the cone structure is less sharply defined. Although frame-dragging remains present, its impact on the shock geometry is partially mitigated by the quantum correction. As $\xi$ increases, the quantity of stuff trapped inside the shock cone drops and the contour lines grow smoother, suggesting a less abrupt transition across the shock. Larger values of $\xi$ thereby lessen strong gravitational focusing and the relative impact of the spin parameter on the shock structure, even in rotating black hole spacetimes.

Finally, the bottom row of Fig.~\ref{color_plots}  presents models IR5 and IR6 corresponding to a rapidly rotating black hole with $a=0.9M$, where the effects of both strong rotation and the quantum correction parameter $\xi$ are most pronounced. The shock cone is strongly bent and azimuthally distorted by frame-dragging associated with rapid rotation, resulting in a pronounced azimuthal asymmetry. A comparatively greater shock opening angle and a less elongated shock cone result from the partial weakening of the spin parameter's influence as $\xi$ grows. As a result, the density of stuff accumulated inside the cone drops. The strong curvature effects characteristic of rapidly rotating Kerr black holes are counteracted by the quantum correction parameter, even if rotation still imprints asymmetry on the flow.

Fig.~\ref{density_angle} illustrates the impact of the black hole spin parameter $a$ and the infrared quantum-gravity correction $\xi$ on the structures formed through the BHL accretion mechanism in the strong-gravity regime. In the left panel, the azimuthal variation of the rest-mass density is shown at a fixed radius $r=2.6M$, i.e., very close to the black hole event horizon, where the effects of rotation and quantum corrections are expected to be maximal. For the GR solutions, namely the Schwarzschild and Kerr spacetimes, the shock cone appears relatively narrow and is characterized by sharp and pronounced density peaks, indicating efficient gravitational focusing and strong compression of the supersonic inflow. In contrast, when infrared quantum-gravity corrections are included, the matter still accretes supersonically toward the black hole, but the compression within the shock cone is significantly reduced, leading to broader azimuthal density profiles and a lower maximum rest-mass density.

A  comparison between the Schwarzschild and Kerr solutions further shows that the black hole spin parameter $a$ induces a systematic azimuthal shift in the location of the maximum density within the shock cone. A redistribution of the post-shock material accompanies this shift and arises directly from frame dragging. On other hand, density profiles of Kerr black holes with asymptotically safe gravity are consistently less peaked and more stretched than those of classical Kerr black holes with the same spin. The modified metric functions, namely the horizon function $\Delta = r^2 - 2Mr + \frac{2M\xi}{r} + a^2$, and the factors $(1-\xi/r^2)$ that show up in the components $g_{tt}$, $g_{t\phi}$, and $g_{\phi\phi}$, are the source of these deviations. These parameters significantly reduce the gravitational potential in the strong-field region of the relativistic
Kerr-like spacetime, which lowers the gravitational focusing efficiency. 
In the strong-field region, these terms effectively weaken the gravitational attraction, thereby reducing the efficiency of gravitational focusing. As a result, the shock compression becomes less effective, the densest part of the post-shock flow spreads over a wider azimuthal region, and the maximum rest-mass density decreases. This behaviour is consistently observed for all considered spin configurations when different values of the quantum-correction parameter $\xi$ are taken into account. of the quantum-correction parameter $\xi$ are taken into account.

The right panel quantifies these morphological changes by showing the dependence of the normalised shock opening angle, $\theta_{\rm sh}/\theta_{\ rmSch,\;Kerr}$, on the quantum parameter $\xi$. For the non-rotating black hole case, increasing $\xi$ leads to an increase in the shock opening angle up to $\sim 4\%$ compared to the Schwarzschild black hole, indicating a substantial reduction in flow collimation. For a rotating black hole (with $a=0.5M$), the deviation from the classical Kerr solution is smaller, at the level of $\sim 2\%$, whereas for a rapidly rotating black hole ($a=0.9M$) the opening angle increases by approximately $\sim 3\%$ compared to the corresponding general relativistic solution. This behaviour reflects the interplay between frame dragging, which tends to azimuthally collimate and distort the shock cone, and the quantum-gravity correction, which counteracts strong gravitational focusing by softening the effective spacetime curvature. In summary, while the black hole spin primarily controls the asymmetry and azimuthal displacement of the shock cone, the quantum parameter $\xi$ governs the collimation strength and density contrast, leading to potentially observable deviations from the classical Schwarzschild and Kerr accretion patterns in the infrared limit of asymptotically safe gravity.

In Fig.\ref {acc_r23}, the top panel shows the time evolution of the mass accretion rate $dM/dt$ measured at $r=2.3M$, which is the closest extraction radius to the compact object considered in our simulations. The accretion rate is evaluated only after the system has reached a quasi-steady state, namely from $t \simeq 3000M$ up to the end of the numerical integration. In this regime, the shock cone has been formed and settled, and the residual temporal variations of $dM/dt$ represent  physical oscillations of the inner accretion flow.  In general, the oscillatory behavior of the numerically obtained mass accretion rate is governed by three main physical mechanisms: (i) the strength of gravitational focusing, which determines how efficiently upstream matter is concentrated into the downstream shock cone, (ii) the degree of post-shock compression inside the cone, which controls the density enhancement and pressure gradients in the inner flow, and (iii) frame dragging effects induced by black hole rotation, which bend streamlines, modify angular momentum transport, and regulate the stability of the shock structure. In the Kerr black hole of asymptotically safe gravity, the lapse and dragging factors differ from the Kerr black hole via a factor of the form $(1-\xi/r^{2})$, which alters the structure of spacetime in the strong-field region and, consequently, changes the dynamical boundary conditions experienced by the accreting fluid. Such effects remain important even in the $a=0$, $\xi=0.6M^{2}$ configuration, which corresponds to a horizonless spacetime, but in this case, the GRHD evolution is performed outside an excised inner region, where the metric remains regular and the fluid dynamics are well defined.

The accretion plateaus (top panel) separate into three bands. The non-rotating black hole exhibits the lowest mean accretion rate $\langle \dot{M} \rangle$, compared to the Schwarzschild solution. The  rotating black hole with $a=0.5M$ occupies an intermediate band close to the corresponding classical Kerr solution, while the rapidly rotating black hole ($a=0.9M$) achieves the highest $\langle \dot{M} \rangle$, again in agreement with the Kerr black holes. The physical expectation that rotation increases inward mass flux through increased relativistic funnelling and a changed streamline topology is consistent with this kind of behavior. Once the downstream shock cone is formed, the frame dragging bends it, redistributes high-density material, and encourages more effective, continuous feeding to the inner extraction sphere.

Concurrently, different values of $\xi$ lead to varied mean accretion levels for a particular spin. When compared to smaller $\xi$, the bigger $\xi$ systematically alters $\langle \dot{M} \rangle$, demonstrating that quantum corrections are not only minute perturbative ripples but can significantly change the overall mass inflow. Physically, this means that $\xi$ changes the way rotation-induced bending and focusing function in the strong field region, in addition to changing the effective dragging forces acting on the falling matter.

The temporal fluctuations suggest that the resulting shock cone sustains oscillatory modes. The strength of these oscillations shows a clear dependence on the black hole rotation parameter $a$, as reflected in the systematic variation of the observed variability with increasing spin.  In particular, the IR2 model corresponding to the naked singularity configuration exhibits a markedly quieter behavior, with strongly suppressed oscillations. A natural physical interpretation is that the excitation and propagation of variability modes change when the strong field region is substantially altered by a big $\xi$. From an observational standpoint, this qualitative distinction is very significant. The IR models show somewhat greater oscillation amplitudes than the  Kerr  black hole for $a=0.5M$. This is in line with the theory that persistent azimuthal shear caused by mild-frame dragging partially stabilizes the cone while permitting low-level modulation driven by shock dynamics. While the IR instances definitely exhibit higher modulation for rapidly spinning black holes ($a=0.9M$), suggesting that quantum corrections might greatly increase variability in the strong frame dragging regime.

The mean mass accretion rate $\langle \dot{M} \rangle$ calculated from the time series in the top panel is shown in the bottom left panel of Fig.\ref {acc_r23}. The corresponding error bars quantify the absolute degree of temporal variations and show the standard deviation of $\dot{M}$. The quantum correction parameter $\xi$ results in non-negligible shifts in $\langle \dot{M} \rangle$ for every spin value. This suggests that, when quantum corrections are taken into account, the mass accretion rates of two compact objects may differ significantly even if they have the same nominal spin. The absolute burstiness of the accretion process is further indicated by the size of the error bars. While the Schwarzschild and low-spin cases exhibit moderate fluctuations, the high-spin configurations generally show smaller absolute dispersion around a higher mean.

Finally, the bottom right panel of Fig.\ref {acc_r23} shows the fractional RMS variability, $\sigma/\langle \dot{M} \rangle$, which measures the strength of fluctuations relative to the average inflow. This quantity is particularly relevant observationally, since in many accretion-powered emission models luminosity variations track, through a transfer function, fluctuations in $\dot{M}$. A larger fractional RMS therefore corresponds to stronger temporal variability and is typically associated with higher QPO amplitudes and enhanced broadband noise. Fig.\ref {acc_r23}, depicts the Schwarzschild and IR1 models exhibit the largest relative variability, indicating a highly active shock cone in the inner region. The IR2 (naked singularity), on the other hand, produces the lowest fractional RMS, suggesting significantly reduced timing noise. The fractional RMS assumes intermediate values for moderate spin.  Overall, these results demonstrate that quantum-corrected geometries do not exhibit a uniform trend but can either enhance or suppress variability depending on $\xi$ and $a$.

Taken together, the bottom panels show that the mean mass accretion rate and the fractional RMS variability provide a two-dimensional behavioral fingerprint of each spacetime model. While black hole rotation primarily sets the accretion rate band, the quantum correction parameter $\xi$ reshapes the variability channel by controlling how coherently and strongly the shock cone dynamics modulate the inflow. Consequently, $\xi$ plays a decisive role in setting both the amplitude of QPO peaks and the level of broadband noise expected in observational diagnostics.

In Fig.\ref {acc_r611}, the same physical quantities discussed in detail for Fig.\ref {acc_r23} are examined at $r=6.11M$, corresponding to a location slightly farther from the strong gravity region. The temporal history of the mass accretion rate is still generally consistent among models when compared to Fig. \ref{acc_r23}, but the oscillation amplitudes are more prominent and, in some cases, amplified in comparison to those seen at smaller radii. This suggests that large-scale flow modulation and shock-cone breathing modes remain dynamically significant away from the immediate vicinity of the black hole.

Although the mass accretion rates corresponding to different spin values remain clearly distinguishable in the top panel of Fig.\ref {acc_r611}, the models sharing the same spin but differing in the quantum correction parameter $\xi$ exhibit only modest differences in their accretion levels. This behaviour suggests that the influence of $\xi$ decreases significantly as one moves away from the immediate vicinity of the black hole, even at radii still associated with strong relativistic effects. Consistently, the bottom panels of Fig.\ref {acc_r611} show that both the mean mass accretion rate and the fractional RMS variability show weaker model-to-model variations with respect to $a$ and $\xi$ at this radius. 

These results agree with theoretical expectations: quantum-corrected gravity effects predominantly modify the innermost strong-field region near the black hole, whereas at larger radii the accretion variability is increasingly governed by shock-cone dynamics and larger-scale flow structures. Consequently, the modulation induced by the shock cone becomes more prominent at $r=6.11M$. In contrast, the direct imprint of quantum corrections on the accretion flow is reduced compared to the behaviour observed closer to the compact object.

Overall, Since both the quantum-correction parameter $\xi$ and the black hole spin parameter $a$ significantly influence the dynamical structure of the shock cone and the associated variability properties, their impact on the resulting physical processes is qualitatively substantial. The black hole spin parameter primarily controls the azimuthal displacement of the shock cone and induces azimuthal asymmetry through frame-dragging effects. In contrast, the quantum-correction parameter $\xi$ symmetrically modifies the strength of gravitational focusing and the compression of matter within the shock cone. As a result, variations in $\xi$ lead to changes in the shock opening angle, post-shock density contrast, and the efficiency with which perturbations are trapped within the post-shock cavity—effects that cannot be reproduced by variations in the spin parameter alone.

%%%%%%%%%%%%%%%%%%%%%%%%%%%%%%%%%%%%%%%%%%%%%%%%%%%%%%%%%%%%%%%%%%%%%%
%%%%%%%%%%%%%%%%%%%%%%%%%%%%%%%%%%%%%%%%%%%%%%%%%%%%%%%%%%%%%%%%%%%%%%
\section{Quantum-Correction Effects on QPOs: Power Spectral Density Analysis}
\label{QPO_analysis}

In the present simulations, the global oscillation modes arise from the presence of a well-defined shock cone formed in the downstream region of the black hole. The shock cone creates a finite post-shock cavity that acts as a resonant region in which perturbations can be trapped and sustained. Radially, these trapped modes oscillate between the inner boundary of the computational domain, where the shock cone is attached,  and the stagnation region of the flow, which is typically located around $r \simeq 26M$ along the downstream axis ($\phi = 0$). At the same time, the flow can oscillate laterally between the boundaries of the shock cone in the azimuthal direction. As a result, pressure and velocity perturbations are confined within the entire post-shock cavity and undergo repeated radial and azimuthal reflections. The characteristic timescales of these oscillations are therefore set by the advection and sound-crossing times across the cavity, leading to coherent global oscillation modes rather than localized, radius-dependent fluctuations. Further details on these modes and their physical origin can be found in Refs.~\cite{Cruz-Osorio:2023wev,Donmez:2024lfi}.

In a black hole characterised by BHL accretion, QPOs naturally arise from the excitation of fundamental downstream shock cone. As is well known from theoretical considerations, quantum-corrected black hole spacetimes modify the structure of spacetime in the strong-field regime and, consequently, alter the properties of the shock cone formed in the accretion flow. Therefore, studying QPOs provides essential insights into the temporal behavior of the accretion flow and offers a promising avenue for testing  black hole models. Motivated by this, we investigate how infrared quantum corrections affect the characteristic QPO frequencies and their amplitudes by performing a detailed PSD analysis. By comparing classical and quantum-corrected models at different extraction radii, we aim to identify robust and physically meaningful signatures of quantum corrections in the variability properties of accreting black holes.

Fig.\ref {QPOs_a00} presents the PSD analysis computed from the temporal oscillations of the mass accretion rate after the system has reached a steady-state, for the Schwarzschild spacetime and the quantum corrected IR1 and IR2 models. The PSD in $r=2.3M$ (left panel) and in $r=6.11M$ (right panel), corresponding to the region near the horizon and a larger-radius extraction region, respectively. We found  similar  characteristic frequencies are observed at both radii thereby the detected QPOs correspond to global oscillation modes of the accretion flow rather than arising from local turbulence or numerical artifacts. In other words, the origin of the QPOs in these systems is physical and can be attributed to the excitation of advective-acoustic modes driven by the shock cone formed around the black hole.

The PSD curves in Fig.\ref {QPOs_a00} further show that modifications of the spacetime geometry can significantly alter both the amplitudes and the frequency distribution of the QPO peaks. When compared to the Schwarzschild  black hole, the IR1 model ($\xi = 0.1M^2$) exhibits enhanced low frequency power at both radii, indicating that large-scale shock cone oscillations are stronger and the accretion flow is more efficiently modulated. This enhancement suggests that QPOs arising in quantum corrected spacetimes with horizons may be more easily observable. while, the IR2 model corresponding to the naked singularity configuration ($\xi = 0.6M^2$) shows a strong suppression of the PSD amplitude, consistent with a dynamically quieter flow in which coherent oscillatory modes are only weakly excited.

The redistribution and shifts of the QPO power relative to the Schwarzschild case are in direct agreement with the changes in the shock cone geometry discussed in Figs.\ref {color_plots} and \ref {density_angle}. As $\xi$ increases, the shock opening angle becomes wider and the density concentration inside the cone is reduced, causing a modification of the effective oscillation cavity. These geometric and dynamical changes provide the physical explanation for the observed variations in QPO amplitudes and frequencies, confirming that the PSD features are genuine signatures of spacetime, induced modifications of the accretion dynamics.

The fundamental modes at $6\,\mathrm{Hz}$ and $15\,\mathrm{Hz}$ can be used as reference frequencies to interpret the observed commensurate structures. The ratios $21.8/15 \simeq 1.45$ and $29/15 \simeq 1.93$ are derived in this instance. The naturally nonlinear character of the flow, temporal stepping effects, and finite grid resolution can all cause these ratios to deviate from the ideal $3{:}2$ and $2{:}1$ values. However, radial and azimuthal modes trapped inside the shock cone can be excited to naturally produce these frequencies.

 The additional peaks observed at $21.8$, $29$, and $67\,\mathrm{Hz}$ are consistent with the nonlinear mode coupling within the post-shock region. The view that the QPOs originate from global shock-cone oscillation modes, whose characteristic frequencies are dictated by the effective advection and sound-crossing periods over the post-shock cavity, is supported by the existence of these quasi-comparative frequency ratios. This further confirms that the detected QPO features are physical manifestations of the coupled flow spacetime system rather than numerical artifacts.

Similarly to Fig.\ref {QPOs_a00}, Fig.\ref {QPOs_a05} presents the PSD analysis for a moderately rotating black hole with spin parameter $a=0.5M$, shown comparatively with the classical Kerr solution. For the quantum corrected Kerr models IR3 and IR4 listed in Table \ref{Initial_data}, corresponding to $\xi=0.1M^2$ and $\xi=0.4M^2$, respectively, the PSDs are computed from the mass accretion rate at $r=2.3M$ and $r=6.11M$. A dominant peak at $\sim14\,\mathrm{Hz}$ is observed in both the classical Kerr and quantum corrected cases. In particular, in the quantum corrected model with correction $\xi=0.1M^2$, the power of the $14\,\mathrm{Hz}$ peak is approximately twice that of the other cases, indicating a significant enhancement of the fundamental mode. In addition to the $14\,\mathrm{Hz}$ feature, our quantum corrected Kerr black holes exhibit additional identifiable PSD peaks at $\sim28$, $42$, $56$, and $70\,\mathrm{Hz}$. The recurrence of the same spectral lines at two physically distinct radii demonstrates that these oscillations arise from global modes of the accretion flow. In other words, the QPOs are set by coherent shock cone and post-shock dynamics, consistent with an advective-acoustic feedback mechanism operating in the downstream cavity, rather than by local numerical effects tied to a particular sampling surface. A similar behavior was already observed in the non-rotating case discussed in Fig.\ref {QPOs_a00}.

Although black hole rotation introduces frame-dragging effects, for  $a=0.5M$  this does not destroy the coherence of the QPO modes. In contrast, the combined effect of moderate rotation and a moderate quantum  parameter $\xi$ leads to clearer and more observationally relevant commensurate frequency structures. Here, the $14\,\mathrm{Hz}$ peak corresponds to the fundamental oscillation mode, while the higher frequency peaks ($\sim28$, $42$, $56$, and $70\,\mathrm{Hz}$) are naturally interpreted as harmonic overtones and nonlinear couplings. The strength of these higher frequency components depends on how sharply the shock compresses the flow and how efficiently perturbations are trapped and fed back within the post-shock region.

When the dominant frequency at $f_0 \simeq 14\,\mathrm{Hz}$ in Fig.\ref {QPOs_a05}  and the accompanying peaks are examined, an almost perfect harmonic sequence emerges. In particular, the ratios $28:14=2$, $42:14=3$, $56:14=4$, and $70:14=5$ form a clear, commensurate ladder of $1{:}2{:}3{:}4{:}5$. Such harmonic structures are in excellent agreement with QPOs observed in astrophysical systems and indicate that the underlying modes are sufficiently coherent and nonlinearly driven.

Compared to the non-rotating case shown in Fig.\ref {QPOs_a00}, the overtones in Fig.\ref {QPOs_a05} appear significantly clearer and more sharply defined. A natural physical explanation is that moderate black hole spin introduces frame dragging induced azimuthal shear, which organizes the downstream post-shock region into a more well-defined resonant cavity. As a result, perturbations are trapped more coherently and efficiently fed back, leading to the emergence of pronounced harmonic overtones. Consequently, the combination of moderate rotation and quantum corrections provides favorable conditions for the excitation of clean, high coherence QPO modes.

Fig.\ref {QPOs_a09} extends the PSD analyses presented in Figs.\ref {QPOs_a00} and \ref {QPOs_a05} to the case of a rapidly rotating black hole with spin parameter $a=0.9M$, comparing the classical Kerr spacetime with the quantum corrected models IR5 ($\xi=0.04M^2$) and IR6 ($\xi=0.095M^2$). By computing the PSDs at $r=2.3M$ and $r=6.11M$, the variability qualities at various radii may be directly compared. PSD amplitudes show a strong radial dependence for $a=0.9M$. The peaks at $r=6.11M$ are much stronger and easier to see, whereas the spectra at $r=2.3M$ are much weaker and more washed out, mimicking low-amplitude broadband transmissions. In particular, for the larger quantum correction parameter (IR6), the detectability of PSD peaks at $r=6.11M$ is substantially enhanced compared to the other cases.

A physically motivated explanation for this behaviour is the strong frame-dragging induced by rapid rotation, which reorganises the inner accretion flow and shifts the most efficient modulation region outward. Near the black-hole horizon, the mass flux through a small extraction sphere becomes comparatively steady, or dominated by incoherent high-shear variability, leading to suppressed PSD amplitudes. While, at larger radii the shock cone breathing and post-shock recirculation remain dynamically active, periodically loading and unloading the accretion flow and producing stronger quasi-periodic modulation.

In $r=6.11M$ in Fig.\ref {QPOs_a09}, prominent PSD peaks are observed in approximately $9$, $17.8$, and $27\,\mathrm{Hz}$. The feature $9\,\mathrm{Hz}$ can be identified as the fundamental frequency, while the higher frequency peaks arise from its nonlinear harmonics, yielding the commensurate ratios $17.8:9  \approx 2{:}1$, $27:9 = 3{:}1$, and $27:17.8  \approx 3{:}2$. The simultaneous presence of harmonic and near $3{:}2$ commensurabilities is a pattern frequently discussed in observational QPO phenomenology, supporting the interpretation that the variability features seen in Fig.\ref {QPOs_a09}  are physical manifestations of global oscillation modes in rapidly rotating, quantum corrected black hole accretion flows.

%%%%%%%%%%%%%%%%%%%%%%%%%%%%%%%%%%%%%%%%%%%%%%%%%%%%%%%%%%%%%%%%%%%%%%
%%%%%%%%%%%%%%%%%%%%%%%%%%%%%%%%%%%%%%%%%%%%%%%%%%%%%%%%%%%%%%%%%%%%%%
\section{Identification of Observable QPO Frequencies from Numerical PSDs}
\label{compareWith_Obs}

In this section, we  investigate the physical validity of the QPO frequencies obtained numerically from the PSD analysis and discuss their consistency with astrophysical observations. Throughout this work, in order to express the numerically calculated frequencies in the SI unit system, we adopt a fiducial stellar-mass black hole with mass $M = 10\,M_\odot$. Consequently, the QPO frequencies identified in the PSD analyses are directly comparable to those observed in Galactic black hole X-ray binaries (XRBs) \cite{Remillard:2006fc}. However, these frequencies can be straightforwardly rescaled, allowing predictions and comparisons to be made for systems hosting massive and supermassive black holes.

As discussed in detail in Section \ref{QPO_analysis}, for both non-rotating and rotating black holes in the infra-red limit of asymptotically safe gravity, the QPO frequencies exhibit systematic variations and the emergence of overtones depending on the black hole spin parameter and the quantum correction term. Here, we explicitly demonstrate how these numerical results are consistent with observations. Since the frequencies identified in Section \ref{QPO_analysis} are obtained by adopting $M = 10\,M_\odot$, the resulting QPO frequencies naturally fall within the range of low- and moderately high-frequency QPOs observed in Galactic XRBs \cite{Donmez:2025piv}.

In particular, the frequencies clustered around $5$--$20~\mathrm{Hz}$ are in one-to-one agreement with the low-frequency QPOs observed in Type-C sources \cite{Casella:2005vy,Ingram:2009vm,Ingram:2019mna}. This class of QPOs is the most frequently detected observationally and is known to arise as a consequence of geometric modulation of the accretion flow and strong variations in its dynamical structure. In this context, the influence of the quantum correction parameter is fully consistent with the modifications of the shock cone structure identified in our numerical simulations.

The appearance of harmonic sequences and near $3\!:\!2$ frequency ratios in the PSD analyses further reveals a strong connection between the simulation results and observed astrophysical sources \cite{Smith:2020npb}. Most importantly, our numerical findings demonstrate that even without invoking epicyclic resonance mechanisms of thin accretion disks, shock-driven global oscillation modes alone can naturally reproduce many of the key timing properties observed in XRBs \cite{Molteni:1995wz}.

On the other hand, the characteristic frequencies obtained in this work are inversely related to the black hole mass. For a numerically identified oscillation mode, the frequency can be rescaled using the relation

\begin{equation}
f(M) = f_{10}\left(\frac{10\,M_\odot}{M}\right),
\label{rescale_Hz}
\end{equation}

\noindent
where $f_{10}$ denotes the QPO frequency calculated from the simulations for a black hole of mass $10\,M_\odot$, and $M$ represents the mass of the black hole to which the frequency is rescaled. Using Eq.\ref{rescale_Hz}, a fundamental QPO frequency of $f_{10} \sim 10$--$20~\mathrm{Hz}$ for stellar-mass black holes corresponds to frequencies of order $10^{-3}$--$10^{-4}~\mathrm{Hz}$ for massive black holes with masses $M \sim 10^{6}$--$10^{7}\,M_\odot$. Accordingly, for such systems the characteristic QPO frequencies lie in the milli-Hertz regime\cite{Donmez:2024luc}. These milli-Hertz frequencies are drelevant to studies of active galactic nuclei and are consistent with observational results \cite{Gierlinski:2008yz,Ren:2023zwf,Reis:2012sz}. Therefore, the global oscillations generated by the post-shock accretion flow, modulated by black hole spin and quantum-corrected spacetime geometry, emerge as a viable physical mechanism capable of explaining QPOs observed across the entire black hole mass spectrum. This mass-scalable nature of QPOs highlights their potential as a unifying probe of strong-field gravity, from X-ray binaries to supermassive black holes \cite{Donmez:2010sx}.

From an observational perspective, our numerical results indicate that variations in the black hole spin, accretion state, and the initial flow parameters of the BHL mechanism may introduce partial degeneracies with the effects induced by the quantum-correction parameter $\xi$. Nevertheless, our findings show that increasing $\xi$ leads to a systematic suppression of QPO peaks in the PSD analysis, a reduction in harmonic richness, and a damping of the coherence of oscillation modes. In contrast, variations in the black hole spin primarily modify the azimuthal dynamics of the accreting matter and the overall efficiency of flow modulation. These differences are reflected in distinct trends of the fractional RMS variability shown in the bottom-right panel of Fig.\ref{acc_r611}. At the same time, the presence and relative strength of harmonic peaks, as well as the coherence properties of the QPOs, further highlight these differences. Such signatures correspond to observable physical variations, indicating that shock-driven timing diagnostics provide an independent channel for probing infrared-modified gravity effects.

Lastly, an observational signature is provided by the dependence of QPO coherence and strength on the quantum correction parameter $\xi$. Strong and coherent QPOs are supported by spacetimes admitting horizons and mild quantum corrections, but oscillations may be greatly repressed and recognizable PSD peaks may be washed out for sufficiently large values of $\xi$. Therefore, precise timing measurements provide a powerful and independent way to assess infrared-modified gravity in the strong-field regime when paired with spectral and imaging diagnostics.

%%%%%%%%%%%%%%%%%%%%%%%%%%%%%%%%%%%%%%%%%%%%%%%%%%%%%%%%%%%%%%%%%%%%%%
%%%%%%%%%%%%%%%%%%%%%%%%%%%%%%%%%%%%%%%%%%%%%%%%%%%%%%%%%%%%%%%%%%%%%%
\section{Conclusion}
\label{concl}
In this study, we demonstrate that the BHL accretion flow around a Kerr-like black hole in asymptotically safe gravity naturally forms a shock cone in the downstream region, i.e., on the side opposite the direction of the incoming matter. The resulting shock cone is shown to depend strongly on both the spin parameter $a$ and the quantum parameter $\xi$. While black hole rotation primarily governs the azimuthal displacement and asymmetry of the shock cone through frame-dragging effects, the quantum correction systematically modifies the strength of gravitational focusing. As a consequence, increasing $\xi$ results in a wider shock opening angle, reduced compression of matter in the post-shock region, and a decrease in the density of material confined within the cone. In extreme cases corresponding to horizonless configurations, the shock cone becomes significantly broader and dynamically quieter. These results indicate that the infrared quantum-gravity correction parameter plays a crucial role in reshaping the accretion flow in the strong-gravity regime.

Global oscillation modes are trapped and sustained within the post-shock region, as evidenced by the time-dependent behaviour of the mass accretion rate. The associated QPOs are identified through power spectral density (PSD) analyses performed after the system has reached a quasi-steady state. Importantly, identical characteristic frequencies are obtained when the PSDs are computed at two physically distinct extraction radii, namely near the compact object and at a larger extraction radius. This radius-independent behaviour demonstrates that the detected QPOs correspond to coherent global oscillation modes of the accretion flow and are not artifacts of a particular sampling location or transient numerical effects. We further note that the global nature of the modes produced by the same numerical framework was previously established through a dedicated time-window analysis (see Fig.~2 of Ref.~\cite{Donmez:2024luc}), where the persistence of the characteristic frequencies over different time intervals was demonstrated. The simulations show the existence of basic modes, harmonic overtones, and near-commensurate frequency ratios like 2:1 and 3:2, depending on the black hole spin and the quantum correction parameters. It is found that totally nonlinear advective-acoustic cycles functioning inside the post-shock cavity are the source of these ratios. The numerical results further the creation of stronger and more distinct oscillation modes is facilitated by modest quantum corrections and moderate black hole rotation. Large quantum correction values, on the other hand, diminish the efficiency with which the post-shock cavity sustains prolonged oscillations by weakening shock compression and suppressing oscillation amplitudes.

Using a stellar-mass black hole $M = 10\,M_\odot$, the numerically calculated QPO frequencies naturally fall within the range of low- and moderately high-frequency QPOs observed in XRBs. The numerical frequencies fall within the characteristic frequency range of Type-C QPOs observed in Galactic X-ray binaries and reproduce several of their key phenomenological features, such as strong coherence and harmonic structure, when scaled to a stellar-mass black hole. The harmonic structures and near $3\!:\!2$ frequency ratios in the simulations are in close agreement with the features reported in observational timing studies. This demonstrates that Shock-driven global oscillation modes can reproduce many of the fundamental QPO characteristics without explicitly invoking epicyclic resonance mechanisms, providing a complementary physical explanation to disk-based models. On the other hand, the inverse mass-scaling relation shows that the frequencies and physical mechanisms identified for XRBs are also applicable to massive black holes. Using this scaling, QPO frequencies in the milli-Hertz range are predicted for massive black holes, in agreement with observational results reported in the literature \cite{Gierlinski:2008yz,Ren:2023zwf,Reis:2012sz}. These findings thus demonstrate QPOs as a physically reliable and mass-scalable probe of strong-field gravity, offering an additional observational channel to test infrared-modified spacetime geometries over the whole black hole mass spectrum.

A full observational disentanglement of black hole spin, accretion conditions, and infrared quantum-gravity effects will ultimately require a multi-observable approach that combines timing diagnostics with spectral, imaging, and independent spin measurements. In contrast, in the present study we focus on shock-driven variability and QPO properties obtained from relativistic hydrodynamic simulations, while recognizing that isolating the impact of the quantum-correction parameter will require breaking parameter degeneracies through complementary observational techniques. In this context, the time-dependent signatures identified in this work—such as variations in QPO peak amplitudes, harmonic structures, coherence properties, and fractional RMS variability that represent independent and physically motivated numerical results that probe the strong-gravity regime. In future studies, we aim to construct a systematic framework by incorporating these observables and exploring a broader parameter space in order to assess their combined diagnostic power. Such an approach will enable a more detailed understanding of the physical consequences of infrared-modified spacetime properties around black holes and provide clearer guidance on how these effects may be detected observationally.

%%%%%%%%%%%%%%%%%%%%%%%%%%%%%%%%%%%%%%%%%%%%%%%%%%%%%%%%%%%%%%%%%%%%%%%%%%%%
\section*{Acknowledgments}
All numerical simulations were performed using the Phoenix High
Performance Computing facility at the American University of the Middle East (AUM), Kuwait.

\section*{Data Availability Statement}
The data sets generated and analyzed during the current study were produced using high-performance computing resources. These data are not publicly available due to their large size and computational nature, but are available from the corresponding author upon reasonable request.

\bibliography{mybibwithouttitle_abbre}

@article{rovelli2008loop,
  title={},
  author={Rovelli, Carlo},
  journal={Living Rev. Relativ.},
  volume={11},
  pages={5},
  year={2008},
  publisher={Springer}
}

@article{azreg2014generating,
  title={Generating rotating regular black hole solutions without complexification},
  author={Azreg-A{\"\i}nou, Mustapha},
  journal={arXiv preprint arXiv:1405.2569},
  year={2014}
}

@article{haroon2018effects,
  title={},
  author={Haroon, Sumarna and Jamil, Mubasher and Lin, Kai and Pavlovic, Petar and Sossich, Marko and Wang, Anzhong},
  journal={Eur. Phys. J. C},
  volume={78},
  pages={519},
  year={2018},
  publisher={Springer}
}

@article{reuter2011quantum,
  title={},
  author={Reuter, M and Tuiran, E},
  journal={Phys. Rev. D},
  volume={83},
  pages={044041},
  year={2011},
  publisher={APS}
}

@article{cai2010black,
  title={},
  author={Cai, Yi-Fu and Easson, Damien A},
  journal={J. Cosmol. Ast. Phys.},
  volume={2010},
  pages={002},
  year={2010},
  publisher={IOP Publishing}
}

@article{pawlowski2024effective,
  title={},
  author={Pawlowski, Jan M and Tr{\"a}nkle, Jan},
  journal={Phys. Rev. D},
  volume={110},
  pages={086011},
  year={2024},
  publisher={APS}
}

@article{niedermaier2006asymptotic,
  title={},
  author={Niedermaier, Max and Reuter, Martin},
  journal={Living Rev. Relati.},
  volume={9},
  pages={5},
  year={2006},
  publisher={Springer}
}

@article{Weinberg1979general,
  title={In General Relativity: An Einstein Centenary Survey},
  author={S. Weinberg, S.W. Hawking, W. Israel},
  journal={Cambridge University Press Cambridge},
  pages={790-831},
  year={1979},
  publisher={Cambridge University Press Cambridge}
}

@article{chiou2015loop,
  title={},
  author={Chiou, Dah-Wei},
  journal={Int. J. Mod. Phys. D},
  volume={24},
  pages={1530005},
  year={2015},
  publisher={World Scientific}
}

@article{ashraf2025orbital,
  title={},
  author={Ashraf, Asifa and Alqahtani, Ali Saeed and Javed, Faisal and Channuie, Phongpichit and Cilli, Arzu and Bouzenada, Abdelmalek and G{\"u}dekli, Ertan and Malik, MY},
  journal={Phys. Dark Universe},
  volume={47},
  pages={101725},
  year={2025},
  publisher={Elsevier}
}

@article{ashraf2025thermal,
  title={},
  author={Ashraf, Asifa and Ditta, Allah and Bouzenada, Abdelmalek and Abd-Elmonem, Assmaa and Abdalla, Nesreen Sirelkhtam Elmki and Atamurotov, Farruh},
  journal={Phys. Dark Universe},
  volume={47},
  pages={101823},
  year={2025},
  publisher={Elsevier}
}

@article{bouzenada2025barrow,
  title={},
  author={Bouzenada, Abdelmalek and Ditta, Allah and Ashraf, Asifa and Maurya, SK and Atamurotov, Farruh and Aslam, Muhammad and Malik, Muhammad Yousaf},
  journal={Nucl. Phys. B},
  pages={116928},
  year={2025},
  publisher={Elsevier}
}

@article{penrose1965gravitational,
  title={},
  author={Penrose, Roger},
  journal={Phys. Rev. Lett.},
  volume={14},
  pages={57},
  year={1965},
  publisher={APS}
}

@article{narzilloev2021dynamics,
  title={},
  author={Narzilloev, Bakhtiyor and Rayimbaev, Javlon and Abdujabbarov, Ahmadjon and Ahmedov, Bobomurat and Bambi, Cosimo},
  journal={Eur. Phys. J. C},
  volume={81},
  pages={269},
  year={2021},
  publisher={Springer}
}

@article{khlopov2014dark,
  title={},
  author={Khlopov, M},
  journal={Int. J. Mod. Phys. A},
  volume={29},
  pages={1443002},
  year={2014},
  publisher={World Scientific}
}

@article{akiyama2019first,
  title={},
  author={Akiyama, Kazunori and Alberdi, Antxon and Alef, Walter and others},
  journal={Astrophys. J. Lett.},
  volume={875},
  pages={L4},
  year={2019},
  publisher={IoP Publishing}
}

@book{rezzolla2013relativistic,
  title={Relativistic hydrodynamics},
  author={Rezzolla, Luciano and Zanotti, Olindo},
  year={2013},
  publisher={OUP Oxford}
}

@article{mustafa2025particle,
  title={},
  author={Mustafa, G and Maurya, SK and Naseer, Tayyab and Cilli, Arzu and G{\"u}dekli, Ertan and Abd-Elmonem, Assmaa and Alhubieshi, Neissrien},
  journal={Nucl. Phys. B},
  volume={1012},
  pages={116812},
  year={2025},
  publisher={Elsevier}
}

@article{kolovs2015quasi,
  title={},
  author={Kolo{\v{s}}, Martin and Stuchl{\'\i}k, Zden{\v{e}}k and Tursunov, Arman},
  journal={Class. Quant. Grav.},
  volume={32},
  pages={165009},
  year={2015},
  publisher={IOP Publishing}
}

@article{stuchlik2013multi,
  title={},
  author={Stuchl{\'\i}k, Zdenek and Kotrlov{\'a}, Andrea and T{\"o}r{\"o}k, Gabriel},
  journal={Astron. Astrophys.},
  volume={552},
  pages={A10},
  year={2013},
  publisher={EDP Sciences}
}

@article{belloni2012high,
  title={},
  author={Belloni, Tomaso M and Sanna, ANDREA and M{\'e}ndez, Mariano},
  journal={Mon. Not. R. Astron. Soc.},
  volume={426},
  pages={1701},
  year={2012},
  publisher={Blackwell Science Ltd Oxford, UK}
}

@inproceedings{rezzolla2004new,
  title={},
  author={Rezzolla, Luciano},
  booktitle={AIP Conf. Proc.},
  volume={714},
  pages={36},
  year={2004},
  organization={American Institute of Physics}
}

@article{remillard2006x,
  title={},
  author={Remillard, Ronald A and McClintock, Jeffrey E},
  journal={Annu. Rev. Astron. Astrophys.},
  volume={44},
  pages={49},
  year={2006},
  publisher={Annual Reviews}
}

@article{wilkins2011determination,
  title={},
  author={Wilkins, DR and Fabian, AC},
  journal={Mon. Not. R. Astron. Soc.},
  volume={414},
  pages={1269},
  year={2011},
  publisher={Blackwell Publishing Ltd Oxford, UK}
}

@article{zare2024shadows,
  title={},
  author={Zare, Soroush and Nieto, Luis M and Feng, Xing-Hui and Dong, Shi-Hai and Hassanabadi, Hassan},
  journal={JCAP},
  volume={2024},
  pages={041},
  year={2024},
  publisher={IOP Publishing}
}

@article{abdujabbarov2016shadow,
  title={},
  author={Abdujabbarov, Ahmadjon and Amir, Muhammed and Ahmedov, Bobomurat and Ghosh, Sushant G},
  journal={Phys. Rev. D},
  volume={93},
  number={10},
  pages={104004},
  year={2016},
  publisher={APS}
}

@article{johannsen2010testing,
  title={},
  author={Johannsen, Tim and Psaltis, Dimitrios},
  journal={Astrophys. J.},
  volume={718},
  pages={446},
  year={2010},
  publisher={IOP Publishing}
}

@article{synge1966escape,
  title={},
  author={Synge, J L},
  journal={Mon. Not. R. Astron. Soc.},
  volume={131},
  pages={463},
  year={1966},
  publisher={Oxford University Press Oxford, UK}
}

@article{johannsen2013photon,
  title={},
  author={Johannsen, Tim},
  journal={Astrophys. J.},
  volume={777},
  pages={170},
  year={2013},
  publisher={IOP Publishing}
}

@article{guo2021universal,
  title={},
  author={Guo, Minyong and Gao, Sijie},
  journal={Phys. Rev. D},
  volume={103},
  pages={104031},
  year={2021},
  publisher={APS}
}

@article{akiyama2022first,
  title={},
  author={Akiyama, Kazunori and Alberdi, Antxon and Alef, Walter and others},
  journal={Astrophys. J. Lett.},
  volume={930},
  pages={L12},
  year={2022},
  publisher={IOP Publishing}
}

@article{akiyama2019firsta,
  title={First M87 event horizon telescope results. V. Physical origin of the asymmetric ring},
  author={Akiyama, Kazunori and Alberdi, Antxon and Alef, Walter and others},
  journal={Astrophys. J. Lett.},
  volume={875},
  pages={L5},
  year={2019},
  publisher={IoP Publishing}
}

@article{akiyama2019firstb,
  title={},
  author={Akiyama, Kazunori and Alberdi, Antxon and Alef, Walter and others},
  journal={Astrophys. J. Lett.},
  volume={875},
  pages={L2},
  year={2019},
  publisher={IOP Publishing}
}

@article{engelhardt2015quantum,
  title={},
  author={Engelhardt, Netta and Wall, Aron C},
  journal={J. High Energy Phys.},
  volume={01},
  pages={73},
  year={2015},
  publisher={Springer}
}

@article{hartman2013time,
  title={},
  author={Hartman, Thomas and Maldacena, Juan},
  journal={J. High Energy Phys.},
  volume={05},
  pages={14},
  year={2013},
  publisher={Springer}
}

@article{barnich2013einstein,
  title={},
  author={Barnich, Glenn and Lambert, Pierre-Henry},
  journal={Phys. Rev. D},
  volume={88},
  pages={103006},
  year={2013},
  publisher={APS}
}

@article{koutrolikos2024just,
  title={},
  author={Koutrolikos, Konstantinos},
  journal={Phys. Rev. D},
  volume={110},
  pages={105010},
  year={2024},
  publisher={APS}
}

@article{dorigoni2024electromagnetic,
  title={},
  author={Dorigoni, Daniele and Duan, Zhihao and Pavarini, Daniele R and Wen, Congkao and Xie, Haitian},
  journal={J. High Energy Phys.},
  volume={2024},
  pages={62},
  year={2024},
  publisher={Springer}
}

@article{brink1977supersymmetric,
  title={},
  author={Brink, Lars and Schwarz, John H and Scherk, Joel},
  journal={Nucl. Phys. B},
  volume={121},
  pages={77},
  year={1977},
  publisher={Elsevier}
}

@article{jackiw1976vacuum,
  title={},
  author={Jackiw, Rf and Rebbi, Claudio},
  journal={Phys. Rev. Lett.},
  volume={37},
  pages={172},
  year={1976},
  publisher={APS}
}

@article{gambini2008black,
  title={},
  author={Gambini, Rodolfo and Pullin, Jorge},
  journal={Phys. Rev. Lett.},
  volume={101},
  pages={161301},
  year={2008},
  publisher={APS}
}

@article{ashtekar2011loop,
  title={},
  author={Ashtekar, Abhay and Singh, Parampreet},
  journal={Class. Quant. Grav.},
  volume={28},
  pages={213001},
  year={2011},
  publisher={IOP Publishing}
}

@article{green1983superstring,
  title={},
  author={Green, Michael B and Schwarz, John H},
  journal={Nucl. Phys. B},
  volume={218},
  pages={43},
  year={1983},
  publisher={Elsevier}
}

@article{hehl1976general,
  title={},
  author={Hehl, Friedrich W and Von der Heyde, Paul and Kerlick, G David and Nester, James M},
  journal={Rev. Mod. Phys.},
  volume={48},
  pages={393},
  year={1976},
  publisher={APS}
}

@article{bojowald2001absence,
  title={},
  author={Bojowald, Martin},
  journal={Phys. Rev. Lett.},
  volume={86},
  pages={5227},
  year={2001},
  publisher={APS}
}

@article{long2019coherent,
  title={},
  author={Long, Gaoping and Lin, Chun-Yen and Ma, Yongge},
  journal={Phys. Rev. D},
  volume={100},
  pages={064065},
  year={2019},
  publisher={APS}
}

@article{bodendorfer2013new,
  title={},
  author={Bodendorfer, Norbert and Thiemann, Thomas and Thurn, Andreas},
  journal={Class. Quant. Grav.},
  volume={30},
  pages={045003},
  year={2013},
  publisher={IOP Publishing}
}

@article{brunnemann2006cosmological,
  title={},
  author={Brunnemann, Johannes and Thiemann, Thomas},
  journal={Class. Quant. Grav.},
  volume={23},
  pages={1395},
  year={2006},
  publisher={IOP Publishing}
}

@book{rovelli2004quantum,
  title={Quantum gravity},
  author={Rovelli, Carlo},
  year={2004},
  publisher={Cambridge university press}
}

@article{hawking1970proceedings,
  title={},
  author={Hawking, S and Penrose, R},
  journal={Proc. R. Soc. Lond. A},
  volume={1970},
  number={314},
  year={1970}
}

@article{misra2020identification,
  title={},
  author={Misra, Ranjeev and Rawat, Divya and Yadav, J S and Jain, Pankaj},
  journal={Astrophys. J. Lett.},
  volume={889},
  pages={L36},
  year={2020},
  publisher={IOP Publishing}
}

@article{chauhan2024spectral,
  title={},
  author={Chauhan, Jaiverdhan and Bharali, Priya and Lohfink, Anne and Abdulghani, Youssef and Davidson, Eric},
  journal={Mon. Not. R. Astron. Soc.},
  volume={527},
  pages={11801},
  year={2024},
  publisher={Oxford University Press}
}

@article{belloni2013discovery,
  title={},
  author={Belloni, Tomaso M and Altamirano, Diego},
  journal={Mon. Not. R. Astron. Soc.},
  volume={432},
  pages={19},
  year={2013},
  publisher={Oxford University Press}
}

@article{liu2021testing,
  title={},
  author={Liu, Honghui and Ji, Long and Bambi, Cosimo and Jain, Pankaj and Misra, Ranjeev and Rawat, Divya and Yadav, JS and Zhang, Yuexin},
  journal={Astrophys. J.},
  volume={909},
  pages={63},
  year={2021},
  publisher={IOP Publishing}
}

@article{motta2016quasi,
  title={Quasi periodic oscillations in black hole binaries},
  author={Motta, Sara E},
  journal={Astron. Nachr.},
  volume={337},
  pages={398},
  year={2016},
  publisher={Wiley Online Library}
}

@article{motta2012discovery,
  title={},
  author={Motta, S and Homan, J and Munoz-Darias, T and Casella, P and Belloni, TM and Hiemstra, B and M{\'e}ndez, M},
  journal={Mon. Not. R. Astron. Soc.},
  volume={427},
  pages={595},
  year={2012},
  publisher={The Royal Astronomical Society}
}

@article{belloni2024fast,
  title={},
  author={Belloni, Tomaso M and M{\'e}ndez, Mariano and Garc{\'\i}a, Federico and Bhattacharya, Dipankar},
  journal={Mon. Not. R. Astron. Soc.},
  volume={527},
  pages={7136},
  year={2024},
  publisher={Oxford University Press}
}

@article{homan2001correlated,
  title={},
  author={Homan, Jeroen and Wijnands, Rudy and van der Klis, Michiel and Belloni, Tomaso and van Paradijs, Jan and Klein-Wolt, Marc and Fender, Rob and Mendez, Mariano},
  journal={Astrophys. J. Suppl. Ser.},
  volume={132},
  pages={377},
  year={2001},
  publisher={IoP Publishing}
}

@article{singh2022low,
  title={},
  author={Singh, Chandra B and Mondal, Santanu and Garofalo, David},
  journal={Mon. Not. R. Astron. Soc.},
  volume={510},
  pages={807},
  year={2022},
  publisher={Oxford University Press}
}

@article{smith2021confrontation,
  title={},
  author={Smith, Krista Lynne and Tandon, Celia R and Wagoner, Robert V},
  journal={Astrophys. J.},
  volume={906},
  pages={92},
  year={2021},
  publisher={IOP Publishing}
}

@article{pasham2025using,
  title={},
  author={Pasham, Dheeraj R and Coughlin, Eric and van Velzen, Sjoert and Hinkle, Jason},
  journal={arXiv preprint arXiv:2502.12078},
  year={2025}
}

@article{pasham2024case,
  title={},
  author={Pasham, Dheeraj R and Tombesi, Francesco and Sukov{\'a}, Petra and Zaja{\v{c}}ek, Michal and others},
  journal={Science Advances},
  volume={10},
  pages={eadj8898},
  year={2024},
  publisher={American Association for the Advancement of Science}
}

@article{pasham2019loud,
  title={},
  author={Pasham, Dheeraj R and Remillard, Ronald A and Fragile, P Chris and others},
  journal={Science},
  volume={363},
  pages={531},
  year={2019},
  publisher={American Association for the Advancement of Science}
}

@article{yousaf2024fuzzy,
  title={},
  author={Yousaf, M and Asad, H and Almutairi, Bander and Hasan, S and Khan, A S},
  journal={Phys. Scr.},
  volume={99},
  pages={115270},
  year={2024},
  publisher={IOP Publishing}
}

@article{javed2023thermal,
  title={},
  author={Javed, Faisal and Mustafa, G and Mumtaz, Saadia and Atamurotov, Farruh},
  journal={Nucl. Phys. B},
  volume={990},
  pages={116180},
  year={2023},
  publisher={Elsevier}
}

@article{zhang2018corrected,
  title={},
  author={Zhang, Ming},
  journal={Nucl. Phys. B},
  volume={935},
  pages={170--182},
  year={2018},
  publisher={Elsevier}
}

@article{pourhassan2017thermodynamics,
  title={},
  author={Pourhassan, B and Kokabi, K and Rangyan, S},
  journal={Gen. Relativ. Gravit.},
  volume={49},
  number={12},
  pages={144},
  year={2017},
  publisher={Springer}
}

@article{gour2003thermal,
  title={},
  author={Gour, Gilad and Medved, AJM},
  journal={Class. Quant. Grav.},
  volume={20},
  number={15},
  pages={3307},
  year={2003},
  publisher={IOP Publishing}
}

@article{hawking1978quantum,
  title={},
  author={Hawking, Stephen W},
  journal={Phys. Rev. D},
  volume={18},
  number={6},
  pages={1747},
  year={1978},
  publisher={APS}
}

@article{harko2010f,
  title={},
  author={Harko, Tiberiu and Lobo, Francisco SN},
  journal={Eur. Phys. J. C},
  volume={70},
  number={1-2},
  pages={373--379},
  year={2010},
  publisher={Springer}
}

@article{harko2011f,
  title={},
  author={Harko, Tiberiu and Lobo, Francisco SN and Nojiri, Shin’ichi and Odintsov, Sergei D},
  journal={Phys. Rev. D},
  volume={84},
  number={2},
  pages={024020},
  year={2011},
  publisher={APS}
}

@article{olmo2019stellar,
  title={},
  author={Olmo, Gonzalo J and Rubiera-Garcia, Diego and Wojnar, Aneta},
  journal={Phys. Rep.},
  volume={876},
  pages={1},
  year={2020}
}

@article{nojiri2017modified,
  title={},
  author={Nojiri, Sh and Odintsov, SD and Oikonomou, VK3683913},
  journal={Phys. Rep.},
  volume={692},
  pages={11},
  year={2017},
  publisher={Elsevier}
}

@article{bamba2014cosmology,
  title={},
  author={Bamba, Kazuharu and Kokusho, Yusuke and Nojiri, Shin'ichi and Shirai, Norihito},
  journal={Class. Quant. Grav.},
  volume={31},
  pages={075016},
  year={2014},
  publisher={IOP Publishing}
}

@article{nojiri2007introduction,
  title={},
  author={Nojiri, S and Odintsov, Sergei D},
  journal={Int. J. Geom. Meth. Mod. Phys.},
  volume={4},
  number={},
  pages={115},
  year={2007},
  publisher={World Scientific}
}

@article{peebles2003cosmological,
  title={},
  author={Peebles, P James E and Ratra, Bharat},
  journal={Rev. Mod. Phys.},
  volume={75},
  pages={559},
  year={2003},
  publisher={APS}
}

@article{perlmutter1999measurements,
  title={},
  author={Perlmutter, Saul and others},
  journal={Astrophys. J.},
  volume={517},
  number={},
  pages={565},
  year={1999},
  publisher={IOP Publishing}
}

@article{riess1998observational,
  title={},
  author={Riess, Adam G and others},
  journal={Astron. J.},
  volume={116},
  pages={1009},
  year={1998},
  publisher={IOP Publishing}
}

@article{riess2007new,
  title={},
  author={Riess, Adam G and others},
  journal={Astrophys. J.},
  volume={659},
  pages={98},
  year={2007},
  publisher={IOP Publishing}
}

@article{ALBADAWI2025102206,
title = {Analytic and Numerical Constraints on QPOs in EHT and XRB Sources Using Quantum-Corrected Black Holes},
journal = {Physics of the Dark Universe},
pages = {102206},
year = {2025},
issn = {2212-6864},
doi = {https://doi.org/10.1016/j.dark.2025.102206},
author = {Ahmad Al-Badawi and Faizuddin Ahmed and Orhan D\"{o}nmez and Fatih Dogan and Behnam Pourhassan and \'{I}zzet Sakallı and Yassine Sekhmani}
}

@Article{Bamba2012,
author="Bamba, Kazuharu
and Capozziello, Salvatore
and Nojiri, Shin'ichi
and Odintsov, Sergei D.",
title="",
journal="Astrophys. Space Sci.",
year="2012",
month="Nov",
day="01",
volume="342",
number="1",
pages="155--228",
issn="1572-946X",
doi="10.1007/s10509-012-1181-8",
url="https://doi.org/10.1007/s10509-012-1181-8"
}

@article{Remillard:2006fc,
    author = "Remillard, Ronald A. and McClintock, Jeffrey E.",
    title = { },
    eprint = "astro-ph/0606352",
    archivePrefix = "arXiv",
    doi = "10.1146/annurev.astro.44.051905.092532",
    journal = "Ann. Rev. Astron. Astrophys.",
    volume = "44",
    pages = "49--92",
    year = "2006"
}

@article{Ingram:2019mna,
    author = "Ingram, Adam and Motta, Sara",
    title = { },
    eprint = "2001.08758",
    archivePrefix = "arXiv",
    primaryClass = "astro-ph.HE",
    doi = "10.1016/j.newar.2020.101524",
    journal = "New Astron. Rev.",
    volume = "85",
    pages = "101524",
    year = "2019"
}

@article{Ingram:2009vm,
    author = "Ingram, Adam and Done, Chris",
    title = { },
    eprint = "0901.1238",
    archivePrefix = "arXiv",
    primaryClass = "astro-ph.SR",
    doi = "10.1111/j.1745-3933.2009.00693.x",
    journal = "Mon. Not. Roy. Astron. Soc.",
    volume = "397",
    pages = "L101",
    year = "2009"
}

@article{Casella:2005vy,
    author = "Casella, Piergiorgio and Belloni, T. and Stella, L.",
    title = { },
    eprint = "astro-ph/0504318",
    archivePrefix = "arXiv",
    doi = "10.1086/431174",
    journal = "Astrophys. J.",
    volume = "629",
    pages = "403--407",
    year = "2005"
}

@article{Donmez:2025piv,
    author = "Donmez, O.",
    title = { },
    doi = "10.1140/epjc/s10052-025-14779-6",
    journal = "Eur. Phys. J. C",
    volume = "85",
    number = "9",
    pages = "1019",
    year = "2025"
}

@article{DONMEZ2026170350,
title = { },
journal = {Annals of Physics},
volume = {486},
pages = {170350},
year = {2026},
issn = {0003-4916},
doi = {https://doi.org/10.1016/j.aop.2026.170350},
author = {Orhan Donmez and Sardor Murodov and Javlon Rayimbaev}
}

@article{Smith:2020npb,
    author = "Smith, Krista Lynne and Tandon, Celia R. and Wagoner, Robert V.",
    title = { },
    eprint = "2011.05346",
    archivePrefix = "arXiv",
    primaryClass = "astro-ph.HE",
    doi = "10.3847/1538-4357/abc9b7",
    journal = "Astrophys. J.",
    volume = "906",
    number = "2",
    pages = "92",
    year = "2021"
}

@article{Ren:2023zwf,
    author = "Ren, Helena X. and Cerruti, Matteo and Sahakyan, Narek",
    title = { },
    doi = "10.22323/1.417.0167",
    journal = "PoS",
    volume = "Gamma2022",
    pages = "167",
    year = "2023"
}

@article{Reis:2012sz,
    author = "Reis, R. C. and Miller, J. M. and Reynolds, M. T. and Gultekin, K. and Maitra, D. and King, A. L. and Strohmayer, T. E.",
    title = { },
    eprint = "1208.1046",
    archivePrefix = "arXiv",
    primaryClass = "astro-ph.CO",
    doi = "10.1126/science.1223940",
    journal = "Science",
    volume = "337",
    pages = "949",
    year = "2012"
}

@article{Gierlinski:2008yz,
    author = "Gierlinski, Marek and Middleton, Matthew and Ward, Martin and Done, Chris",
    title = { },
    eprint = "0807.1899",
    archivePrefix = "arXiv",
    primaryClass = "astro-ph",
    month = "7",
    year = "2008"
}

@article{Molteni:1995wz,
    author = "Molteni, Diego M. and Sponholz, H. and Chakrabarti, Sandip K.",
    title = { },
    eprint = "astro-ph/9508022",
    archivePrefix = "arXiv",
    doi = "10.1086/176775",
    journal = "Astrophys. J.",
    volume = "457",
    pages = "805",
    year = "1996"
}

@article{Donmez:2010sx,
    author = "Donmez, Orhan and Zanotti, Olindo and Rezzolla, Luciano",
    title = { },
    eprint = "1010.1739",
    archivePrefix = "arXiv",
    primaryClass = "astro-ph.HE",
    doi = "10.1111/j.1365-2966.2010.18003.x",
    journal = "Mon. Not. Roy. Astron. Soc.",
    volume = "412",
    pages = "1659--1668",
    year = "2011"
}

@article{Donmez:2024luc,
    author = "Donmez, Orhan and Dogan, Fatih",
    title = { },
    eprint = "2407.01478",
    archivePrefix = "arXiv",
    primaryClass = "gr-qc",
    doi = "10.1016/j.dark.2024.101718",
    journal = "Phys. Dark Univ.",
    volume = "46",
    pages = "101718",
    year = "2024"
}

@article{Mustafa:2025mkc,
    author = "Mustafa, G. and Ghosh, Sushant G. and Donmez, Orhan and Maurya, S. K. and Orzuev, Shakhzod and Atamurotov, Farruh",
    title = { },
    eprint = "2506.16405",
    archivePrefix = "arXiv",
    primaryClass = "gr-qc",
    doi = "10.1088/1475-7516/2025/10/068",
    journal = "JCAP",
    volume = "10",
    pages = "068",
    year = "2025"
}

@article{Donmez:2004ss,
    author = "Donmez, Orhan",
    title = { },
    eprint = "gr-qc/0406073",
    archivePrefix = "arXiv",
    doi = "10.1023/B:ASTR.0000044610.53714.95",
    journal = "Astrophys. Space Sci.",
    volume = "293",
    pages = "323--354",
    year = "2004"
}

@article{Donmez:2005mc,
    author = "Donmez, Orhan",
    title = { },
    eprint = "gr-qc/0512104",
    archivePrefix = "arXiv",
    doi = "10.1016/j.amc.2006.01.031",
    journal = "Appl. Math. Comput.",
    volume = "181",
    pages = "256--270",
    year = "2006"
}

@article{Donmez:2014bza,
    author = {D{\"o}nmez, Orhan},
    title = { },
    doi = "10.1142/S0218271814500503",
    journal = "Int. J. Mod. Phys. D",
    volume = "23",
    pages = "1450050",
    year = "2014"
}

@article{2012MNRAS.426.1533D,
       author = {{D{\"o}nmez}, O.},
        title = { },
      journal = {MNRAS},
         year = 2012,
        month = oct,
       volume = {426},
       number = {2},
        pages = {1533-1545},
          doi = {10.1111/j.1365-2966.2012.21616.x}
}

@article{Donmez:2024lfi,
    author = "Donmez, Orhan",
    title = { },
    eprint = "2402.16707",
    archivePrefix = "arXiv",
    primaryClass = "astro-ph.HE",
    doi = "10.1088/1475-7516/2024/09/006",
    journal = "JCAP",
    volume = "09",
    pages = "006",
    year = "2024"
}

@article{Cruz-Osorio:2023wev,
    author = "Cruz-Osorio, Alejandro and Rezzolla, Luciano and Lora-Clavijo, Fabio Duvan and Font, Jos{\'e} Antonio and Herdeiro, Carlos and Radu, Eugen",
    title = { },
    eprint = "2301.06564",
    archivePrefix = "arXiv",
    primaryClass = "astro-ph.HE",
    doi = "10.1088/1475-7516/2023/08/057",
    journal = "JCAP",
    volume = "08",
    pages = "057",
    year = "2023"
}

\end{document}